\documentclass[english]{article}
\usepackage[T1]{fontenc}
\usepackage[latin9]{inputenc}
\usepackage{geometry}
\usepackage{amsmath}
\usepackage{amsthm}
\usepackage{amssymb}
\usepackage{graphicx}

\makeatletter
\theoremstyle{remark}
\newtheorem*{rem*}{\protect\remarkname}
\theoremstyle{plain}

\theoremstyle{plain}
\newtheorem*{assumption*}{\protect\assumptionname}
\theoremstyle{plain}

\theoremstyle{plain}
\newtheorem*{algorithm*}{\protect\algorithmname}

\@ifundefined{showcaptionsetup}{}{%
 \PassOptionsToPackage{caption=false}{subfig}}
\usepackage{subfig}
\makeatother

\usepackage{babel}
\providecommand{\algorithmname}{Algorithm}
\providecommand{\remarkname}{Remark}
\providecommand{\assumptionname}{Assumption}
\providecommand{\corollaryname}{Corollary}

\providecommand{\theoremname}{Theorem}

\newcommand{\address}[1]{
	\par {\raggedright #1
	\vspace{.5em}
	\noindent\par}
}

\begin{document}
\title{Anomalous Statistics and Large Deviations\\ of Turbulent Water Waves past a Step
}
\author{Di Qi\textsuperscript{a} and Eric Vanden-Eijnden\textsuperscript{b}}
\maketitle
\address{\textsuperscript{a}Department of Mathematics, Purdue University, West Lafayette, IN 47907; \textsuperscript{b}Department of Mathematics, Courant Institute of Mathematical Sciences, New York University, New York, NY 10012}

\begin{abstract}
A computational strategy based on large deviation theory (LDT) is used to study the anomalous statistical features of turbulent surface waves propagating past an abrupt depth change created via a step in the bottom topography. The dynamics of the outgoing waves past the step are modeled using the truncated Korteweg-de Vries (TKdV) equation with random initial conditions at the step drawn from the system's Gibbs invariant measure of the incoming waves. Within the LDT framework, the probability distributions of the wave height can be obtained via the solution of a deterministic optimization problem.
Detailed numerical tests show that this approach accurately captures the non-Gaussian features of the wave height distributions, in particular their asymmetric tails leading to high skewness. These calculations also give the spatio-temporal pattern of the anomalous waves most responsible for these non-Gaussian features.  The strategy shows potential for a general class of nonlinear Hamiltonian systems with highly non-Gaussian statistics.
\end{abstract}

\section{Introduction}
\label{sec:intro}

Recent laboratory experiments \cite{tru2012lab,bolles2019anomalous} with unidirectional water waves propagating in a tank with a variable bottom have shown that a brutal depth change produces outgoing waves with anomalous statistical features. In particular, the wave height distribution can become highly non-Gaussian past this step, shortening the tail of the distribution of large negative amplitudes and fattening the one of large positive amplitudes \cite{viotti2014extreme,qi2016predicting,qi2018predicting}. This mechanism has been invoked as a possible explanation for the apparition of rogue waves in the shallow-to-moderate depth regime where the classical Benjamin-Feir (BF) instability thought to be at play for rogue waves in deep water is absent~\cite{onorato2013rogue,Adcock2014ThePO,cousins2015unsteady,onorato2016twenty,dematteis2019experimental}. In Refs.~\cite{majda2019statistical,majda2019transition} a statistical model based on the truncated Kortweg-de Vries (TKdV) equation was shown to correctly describe the anomalous wave statistics past the step observed in the experiments. In particular this model captures the statistical transition from a near-Gaussian state before the step to a  non-Gaussian state highly skewed towards large wave amplitudes past this step \cite{moore2020anomalous,sun2020rigorous}.

The aim of the present paper is to revisit the predictions from the TKdV model using the approach proposed in Refs. \cite{dematteis2018rogue,dematteis2019extreme}, based on large deviation theory (LDT) \cite{varadhan1966asymptotic,freidlin1998random}. Specifically, we consider the TKdV equation with random initial data at the step and investigate the statistics of the solution past this step. The initial conditions are assumed to be drawn either from a Gaussian initial distribution or, more realistically, from the non-Gaussian Gibbs distribution \cite{bajars2013weakly} capturing the non-equilibrium statistical steady state of the solution of the TKdV equation incoming the step. The statistics of the solution outgoing the step can in principle be calculated by brute-force Monte-Carlo (MC) sampling: this amounts to generating many initial conditions at the step and following their evolution past the step. This approach is straightforward, but it has the disadvantage of being inefficient to capture rare events involving e.g. the waves with large positive amplitudes that fatten the right tail of their distribution. When it applies, LDT provides us with an alternative way to perform these calculations by identifying the wave most likely to reach a certain height.  This most likely wave is the minimizer of an action, meaning that MC sampling is replaced by the solution of a deterministic optimization problem. The solution to this problem not only gives an estimate of the probability distribution of the wave height, but it also characterizes the mechanism by which such waves with high positive or negative amplitudes occur. The LDT estimate for the probability can also be refined by calculating a prefactor. 

Here, we formulate the LDT approach for the TKdV equations with random initial data and use numerical tests to confirm the accuracy of the prediction of this approach for the wave amplitude distribution. Compared to the method used in Refs. \cite{dematteis2019extreme,grafke2019numerical,tong2020extreme}, we also consider  non-Gaussian initial data drawn form a Gibbs distribution, which introduces additional difficulties.  We show that LDT correctly captures the asymmetric tails in the distribution as well as the extreme waves observed in direct numerical solutions. The approach thereby provides us with an efficient way to bypass extensive MC sampling simulations and identify directly the waves with large positive or negative heights that dominate the statistics.

The remainder of this paper is organized as follows: In Sec.~\ref{sec:model_formulation} we formulate the statistical model  using the TKdV equation with different initial distributions. In Sec.~\ref{sec:Strategies-for-predicting} we describe the LDT strategies with refined prefactor calibration to predict the probability of extreme events with Gaussian and non-Gaussian initial states. Detailed numerical simulations are performed to confirm the LDT prediction for Gaussian and non-Gaussian initial distributions in Secs. \ref{sec:Numerics-to-compute} and \ref{sec:Numerical-tests-Gibbs} respectively. We close the paper with concluding remarks in Sec.~\ref{sec:summary}.

\section{A simplified statistical framework for modeling surface wave turbulence}\label{sec:model_formulation}

Here we review the statistical model based on the TKdV equation which has been shown to correctly describes the downstream wave state past a step~\cite{bolles2019anomalous,moore2020anomalous}. For more details about the model derivation and rescaling model parameters for fitting experimental setup  we refer the readers to Ref. \cite{moore2020anomalous}. 

\subsection{Surface wave turbulence with water depth dependence\label{subsec:Surface-wave-turbulence}}

The KdV equation is a standard model to describe the anomalous behavior of surface water waves traveling
in a one-dimensional channel \cite{johnson1997modern}.
Following Refs.~\cite{majda2019statistical,moore2020anomalous}, we truncate the equation and introduce and a water depth dependence $D$ to model the turbulent wave dynamics outgoing a step, arriving at the TKdV equation
\begin{equation}
\frac{\partial u_{\Lambda}}{\partial t}+\frac{1}{2}C_{3}D^{-\frac{3}{2}}\frac{\partial}{\partial x}\mathcal{P}_{\Lambda}\left(u_{\Lambda}^{2}\right)+C_{2}D^{\frac{1}{2}}\frac{\partial^{3}u_{\Lambda}}{\partial x^{3}}=0,\qquad0\leq t\leq T.\label{eq:model}
\end{equation}
Here the state variable describing surface wave displacement, $u_{\Lambda}\left(t,x\right)=\mathcal{P}_{\Lambda}u=\sum_{0\leq\left|k\right|\leq\Lambda}\hat{u}_{k}\left(t\right)e^{ikx}$, is Galerkin truncated to a largest wavenumber $\Lambda$. Fourier modes are adopted based on the periodic boundary condition for the normalized spatial coordinate $x\in\left[-\pi,\pi\right]$.  The model coefficient $C_{3}$ characterizes the strength of the nonlinear coupling, while $C_{2}$ gives the strength of the linear dispersive effect. 

An abrupt depth change (ADC) consistent with the lab experiment \cite{bolles2019anomalous} is modeled by a sudden shift in the water depth $D$ at the initial time $t=0$. The incoming waves with normalized water depth $D_{-}=1$ are assumed to have reached equilibrium before the initial time, and used as random initial data in~\eqref{eq:model}, as described next in Sec.~\ref{sec:incoming}. The TKdV equation~\eqref{eq:model} then describes the downstream wave state with a shallower water
depth $D < 1$ emerging from these initial data after the ADC.

\subsection{Statistical characterization of the incoming wave}
\label{sec:incoming}

The TKdV equation (\ref{eq:model}) can also be written as
\begin{equation}
\partial_{t}u_{\Lambda}=\partial_{x}\frac{\delta\mathcal{H}}{\delta u_{\Lambda}},\qquad\mathcal{H}\left(u_{\Lambda}\right)=C_{3}D^{-\frac{3}{2}}H_{3}\left(u_{\Lambda}\right)-C_{2}D^{\frac{1}{2}}H_{2}\left(u_{\Lambda}\right)\label{eq:hamil}
\end{equation}
where $\mathcal{H}\left(u_{\Lambda}\right)$ is the Hamiltonian with a cubic component $H_{3}\left(u\right)=\frac{1}{6}\int_{-\pi}^{\pi}u^{3}dx$ and a quadratic component $H_{2}\left(u\right)=\frac{1}{2}\int_{-\pi}^{\pi}u_{x}^{2}dx$. Besides $\mathcal{H}\left(u_{\Lambda}\right)$, there are two additional invariant quantities during the
evolution, that is, the momentum $\mathcal{M}\left(u_{\Lambda}\right)=\int_{-\pi}^{\pi}u_{\Lambda}dx$ and the total energy
$\mathcal{E}\left(u_{\Lambda}\right)=\frac{1}{2}\int_{-\pi}^{\pi}u_{\Lambda}^{2}dx$.

If the incoming wave has been propagating for a long time and has reached a statistical steady state, its invariant measure can be derived using tools from statistical mechanics based on the Hamiltonian structure of the tKdV equation \cite{abramov2003hamiltonian,bajars2013weakly}.
Following Ref. \cite{majda2019statistical}, we will model this invariant measure as the Gibbs distribution $\mathcal{G}_{-}(u_{\Lambda})$ 
\begin{equation}
\mathcal{G}_{-}\left(u_{\Lambda}\right)=Z_{\theta}^{-1}\exp\left[-\theta \mathcal{H}_-\left(u_{\Lambda}\right)\right],\qquad \text{if} \ \ \  \mathcal{E}\left(u_{\Lambda}\right)\le E_{0},\label{eq:gibbs_invar}
\end{equation}
and $\mathcal{G}_{-}\left(u_{\Lambda}\right) = 0$ if 
$\mathcal{E}\left(u_{\Lambda}\right)> E_{0}$. Here $Z_{\theta}$ is a normalization factor,
$\mathcal{H}_-\left(u_{\Lambda}\right)$ is the Hamiltonian \eqref{eq:hamil} with the upstream water depth $D_{-}=1$, and $\theta$ is the `inverse temperature' determined by the upstream solution statistic. The cap in energy $\mathcal{E}$ is required to guarantee that the probability distribution~\eqref{eq:gibbs_invar} is normalizable (that is, $Z_{\theta}<\infty$) since $\mathcal{H}_{-}$ is bounded on the ball $\mathcal{E}\left(u_{\Lambda}\right)\le E_{0}$. The Gibbs distribution permits one to predict the statistics of the incoming waves without having to run expensive direct simulations of the full model. These predictions are also in agreement with the data from laboratory experiments \cite{moore2020anomalous}.

For comparison,  we will also consider the case where the  initial state is drawn from a Gaussian distribution with an energy spectrum $\left\{ R_{k}\right\} _{k=0}^{\Lambda}$ characterizing the variance of each mode, that is,
\begin{equation}
u_{0}\left(x\right)=u_{\Lambda}\left(0,x\right)=\sum_{\left|k\right|\leq\Lambda}\hat{u}_{k}\left(0\right)e^{ikx},\qquad\hat{u}_{k}\left(0\right)=\sqrt{R_{k}}\,\hat{\xi}_{k},\qquad \hat{\xi}_{k}\overset{\mathrm{i.i.d.}}{\sim}\mathcal{N}\left(0,1\right).\label{eq:init_gau}
\end{equation}
where the initial energy spectrum $R_{k}$ can be defined from the variance of each mode $\hat{u}_{k}$  as a Gaussian fit of the sampled Gibbs distribution \eqref{eq:gibbs_invar}.
Using the Gaussian initial state in (\ref{eq:init_gau}) is a valid assumption when the  Gibbs distribution (\ref{eq:gibbs_invar}) is close to a Gaussian distribution in the incoming flow state \cite{majda2019statistical}. Experimental observations \cite{bolles2019anomalous} support this assumption of near-Gaussian statistics in the incoming surface wave distribution before the ADC. However, this prediction is only valid in the core of the distribution and we will see below that the difference in the tail between the Gibbs and the Gaussian distributions will have a nontrivial impact on the statistics of the outgoing waves.

\subsection{Statistical characterization of the outgoing wave}
\label{sec:outgoing}

Below we will look for downstream solutions of the TKdV equation~\eqref{eq:model} for $t>0$ after the ADC  using random initial condition at $t=0$ that are either drawn from the Gibbs distribution~\eqref{eq:gibbs_invar} or Gaussian generated using~\eqref{eq:init_gau}. These solutions are random through the uncertainty in their initial conditions, and we seek to characterize their statistical properties. Note in particular that we cannot expect the transient statistics of the outgoing waves to be described by the Gibbs distribution for $t>0$ since the depth has changed, i.e. we use $D<D_-=1$ in~\eqref{eq:model}.  We will mainly focus on the positive and negative surface wave displacements evaluated at location $x=x_{c}$ and  time $t=T>0$,
that is for $z>0$, we will estimate
\begin{equation}
\begin{aligned}P_{T}^{+}\left(z\right)= & \:\mathbb{P}\left\{ u_{\Lambda}\left(T,x_{c}\right)\geq z\right\} ,\\
P_{T}^{-}\left(z\right)= & \:\mathbb{P}\left\{ u_{\Lambda}\left(T,x_{c}\right)\leq-z\right\}.
\end{aligned}
\label{eq:probs}
\end{equation}

\section{Predicting asymmetric tails using large deviation principle\label{sec:Strategies-for-predicting}}

Next, we describe the strategy to efficiently compute
the asymmetric tails in probability distributions for the TKdV equation (\ref{eq:model}) with either Gaussian and non-Gaussian initial states using the large deviation framework proposed in Refs. \cite{dematteis2019extreme,tong2020extreme}.

\subsection{Large deviation principle with Gaussian initial state}

The probabilities $P_{T}^{\pm}\left(z\right)$ in \eqref{eq:probs} from Gaussian initial data can be computed from integration about the initial coefficients according to the admissible event set $\Omega_{T}^{\pm}\left(z\right)$ given extreme values at the final measurement time $T$ as
\begin{equation}
P_{T}^{\pm}\left(z\right)=Z_{\pm}^{-1}\int_{\Omega_{T}^{\pm}\left(z\right)}\exp\left(-\frac{1}{2}\left\Vert \xi\right\Vert ^{2}\right)d\xi,\qquad \Omega_{T}^{\pm}\left(z\right)=\left\{ \xi: \pm u_{\Lambda}\left(T,x_{c};\xi\right)\geq z >0\right\},
\label{eq:integration}
\end{equation}
where we use the shorthand $\xi\in\mathbb{R}^{2\Lambda+1}$ to collectively denote  the real and imaginary parts of all the spectral modes,  $\left\Vert \xi\right\Vert ^{2}=\sum_{k}|\hat{\xi}_{k}|^{2}$ is the $L_{2}$-norm of the initial coefficients, and $Z_{\pm}$ are normalization factors. 

Performing the integration in (\ref{eq:integration}) is difficult because  $\Omega_{T}^{\pm}\left(z\right)$  are complicated sets involving  the solution $u_{\Lambda}\left(T,x\right)$ after propagation by the nonlinear TKdV equation (\ref{eq:model}) from $t=0$ till $t=T$. LDT allows us to bypass this calculation in situation where the integrals in~\eqref{eq:integration} are dominated by one single extreme trajectory that maximizes the integrand, leading to the following Laplace estimates \cite{nocedal2006numerical}
\begin{equation}
P_{T}^{\pm}\left(z\right)\asymp \exp\left[-I_{T}^{\pm}\left(z\right)\right],\qquad
I_{T}^{\pm}\left(z\right)= \min_{\xi\in\Omega_{T}^{\pm}\left(z\right)}\frac{1}{2}\sum_{\left|k\right|\leq\Lambda}|\hat{\xi}_{k}|^{2},
\label{eq:LDT}
\end{equation}
where the symbol `$\asymp$' in (\ref{eq:LDT}) denotes to exponential convergence, i.e.
\begin{equation}
\lim_{z\rightarrow\infty}\frac{\log P_{T}^{\pm}\left(z\right)}{I_{T}^{\pm}\left(z\right)}=-1.\label{eq:limit}
\end{equation}
In practice, the minimization problem in~\eqref{eq:LDT}  can be calculated by introducing a Lagrangian multiplier $\lambda>0$, leading to
\begin{equation}
\xi_\pm^{*}\left(\lambda\right)=\underset{\xi\in\Omega}{\arg\min}\left\{ \tfrac12\|\xi\|^2\mp\lambda u_{\Lambda}\left(T,x_{c};\xi\right)\right\} ,\label{eq:optim_gau}
\end{equation}
and the value of the amplitude can then be recovered from  $z=\pm u_{\Lambda}\left(T,x_{c};\xi^*_\pm(\lambda)\right)$. This gives a parametric relation between $P_{T}^{\pm}\left(z\right)$ and $z$ as a function of $\lambda$.

\subsubsection*{A refined approximation with a higher-order correction}

As shown in Ref. \cite{tong2020extreme}, the probability
distribution can be approximated more accurately by introducing a higher-order correction
\begin{equation}
P_{T}\left(z\right)\approx C\left|\xi^{*}\left(z\right)\right|^{-1}\exp\left[-I_{T}\left(z\right)\right],\label{eq:expansion_gau}
\end{equation} 
where $C$ is a normalization constant.
Since the TKdV model (\ref{eq:model}) conserves the total energy, the LDT initial state requires larger total energy $\mathcal{E}=|\xi^{*}\left(z\right)|^{2}\rightarrow\infty$ as a larger extreme valuee is reached as $z\rightarrow\infty$. Thus the second correction term in (\ref{eq:expansion_gau}) shows a slower growth rate compared with the leading term $I_{T}\left(z\right)$ as a refined
correction to the probability.

\subsection{Large deviation principle for non-Gaussian Gibbs initial state}\label{subsec:gibbs}

If we consider the statistics of the solutions to the TKdV equation~\eqref{eq:model}  with initial conditions drawn from the Gibbs distribution~\eqref{eq:gibbs_invar}, we are interested in computing
\begin{equation}
    P^\pm _{T}\left(z\right)=Z_{\theta}^{-1}\int_{\Omega^\pm_{T}\left(z\right)}\exp\left[-\theta \mathcal{H}_-(u_0) \right]du_{0},
\end{equation}
where $u_{0}\in\Omega\subseteq\mathbb{R}^{2\Lambda+1}$ and  $\Omega^{\pm}_{T}\left(z\right)=\left\{ u_{0}: \pm u_{\Lambda}\left(T,x_{c};u_{0}\right)=z,\:\mathcal{E}\left(u_0\right)\leq E_0\right\} $. Consistent with the LDT approach, we perform this integral using Laplace method, which leads to the following minimization problem with two imposed Lagrangian multipliers $\lambda$ and $\mu$
\begin{equation}
u_{0}^{*,\pm}\left(\lambda,\mu\right)=\underset{u_{0}\in\Omega}{\arg\min}\left\{ \theta\mathcal{H}_-\left(u_{0}\right)+\mu\mathcal{E}\left(u_{0}\right)\mp\lambda u_{\Lambda}\left(T,x_{c};u_{0}\right)\right\} .\label{eq:optim_gibbs}
\end{equation}
In this expression the pair of parameters $\left(\lambda,\mu\right)$ determines the LDT values of extreme event and corresponding energy $\left(z,E\right)$ via $z=\pm u_{\Lambda}\left(T,x_{c};u_0^{*,\pm}(\lambda,\mu)\right)$ and $E_0 = \mathcal{E}_0\left(u_{\Lambda}\left(T,x_{c};u_0^{*,\pm}(\lambda,\mu)\right)\right)$. By changing the value of $(\lambda,\mu)$, we can therefore determine the range of permitted total energy $E$ that leads to an extreme state of amplitude $z$. LDT then predicts the log-asymptotic behavior of the probability with the corresponding action functional defined in this case by the Hamiltonian for $z\gg1$ via
\begin{equation}
P^\pm_{T}\left(z\right)\asymp\exp\left(-\theta \mathcal{H}_-\left(u_{0}^{*,\pm}\right)\right)=\exp\left(-\theta C_{3}D_-^{-\frac{3}{2}}H_{3}\left(u_{0}^{*,\pm}\right)+\theta C_{2}D_-^{\frac{1}{2}}H_{2}\left(u_{0}^{*,\pm}\right)\right).\label{eq:LDT_gibbs-1}
\end{equation}

\section{Numerical results from the LDT prediction for Gaussian initial data\label{sec:Numerics-to-compute}}

In this section, we check the prediction skill of the LDT strategy described in Sec.~\ref{sec:Strategies-for-predicting} in capturing the tails of probability distributions  under Gaussian initial data. For the turbulent waves modeled by the TKdV equation (\ref{eq:model}), we are particularly interested in the skewed PDF which implies asymmetry in the positive and negative extreme values that are observed in laboratory experiments \cite{bolles2019anomalous}.
The solutions from the LDT approach characterizes the trajectories by which such extreme values arise, thus offering a useful tool to understand the physical mechanism behind such phenomena. 

\subsection{Numerical setup}

\begin{figure}
\includegraphics[scale=0.40]{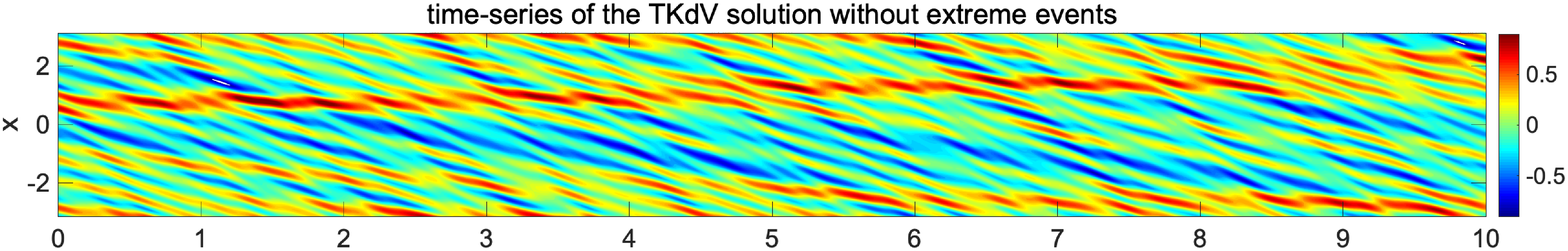}
\includegraphics[scale=0.40]{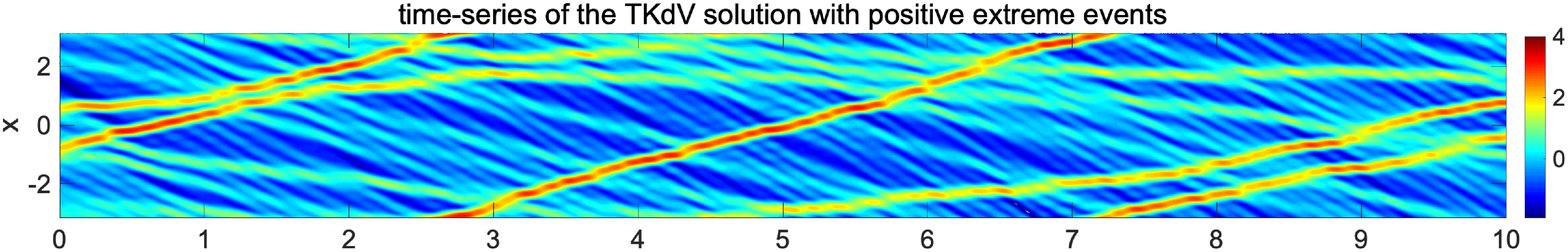}
\includegraphics[scale=0.40]{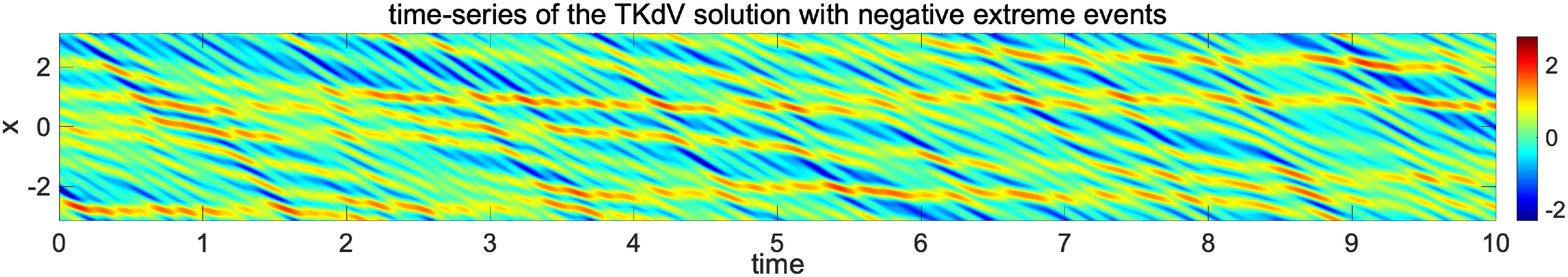}

\caption{Solutions of the TKdV equation in three typical trajectories selected from the ensemble simulation. The trajectories in the three regimes without extreme events (upper), with positive extreme values (middle), and with negative extreme values (lower) are compared. Notice the different ranges of the color scale in the three regimes.\label{fig:Solutions-of-tKdV}}
\end{figure}

The TKdV equation (\ref{eq:model}) for the wave statistical transition is solved with the initial state sampled from the incoming flow statistics approximated by a Gaussian distribution (\ref{eq:init_gau}) with spectrum $\mathbb{E}\left|\hat{u}_{k}\right|^{2}=R_{k}$. We employ a pseudo-spectral scheme to solve the TKdV equation and a 4th-order midpoint symplectic scheme \cite{mclachlan1993symplectic} for the time integration so that the conservation property is preserved. The true model statistics are estimated from a direct Monte-Carlo simulation of the original TKdV equation (\ref{eq:model}) with a large ensemble
of $1\times10^{6}$  initial samples in order to capture the exponential tails of the PDFs with accuracy. We pick the shallower water depth $D=0.24$ and model coefficients $C_{2}=0.0236$, $C_{3}=0.2312$, and model truncation size as $\Lambda=16$. The detailed numerical arrangement of the equation from the rescaled experimental configuration is summarized in Refs. \cite{moore2020anomalous,majda2019statistical}. Typical solutions of the TKdV equation are displayed in Figure \ref{fig:Solutions-of-tKdV}, which shows realizations of the TKdV model solutions with and without extreme values. 

The different statistics and skewness in the PDFs are produced from the initial energy spectra $R_{k}$ for the Gaussian initial data. We focus on two typical cases of initial spectra: \emph{case
I} with a decaying energy spectrum; and \emph{case II} with  equipartition
energy (see the first column of Figure \ref{fig:LDT-predictions-CDF}).
The two initial spectra are consistent with the sampled variances from the Gibbs distribution \eqref{eq:gibbs_invar}, and agree with the spectra observed in the laboratory experiments \cite{bolles2019anomalous}. The detailed numerical strategy to compute the LDT solutions is described in the Appendix.

\subsection{Predicting probability distributions in positive and negative extreme values}

\subsubsection{Prediction of tails in probability distributions with different initial spectra}

\begin{figure}
\subfloat[Case I: decaying initial energy spectrum]{\includegraphics[scale=0.29]{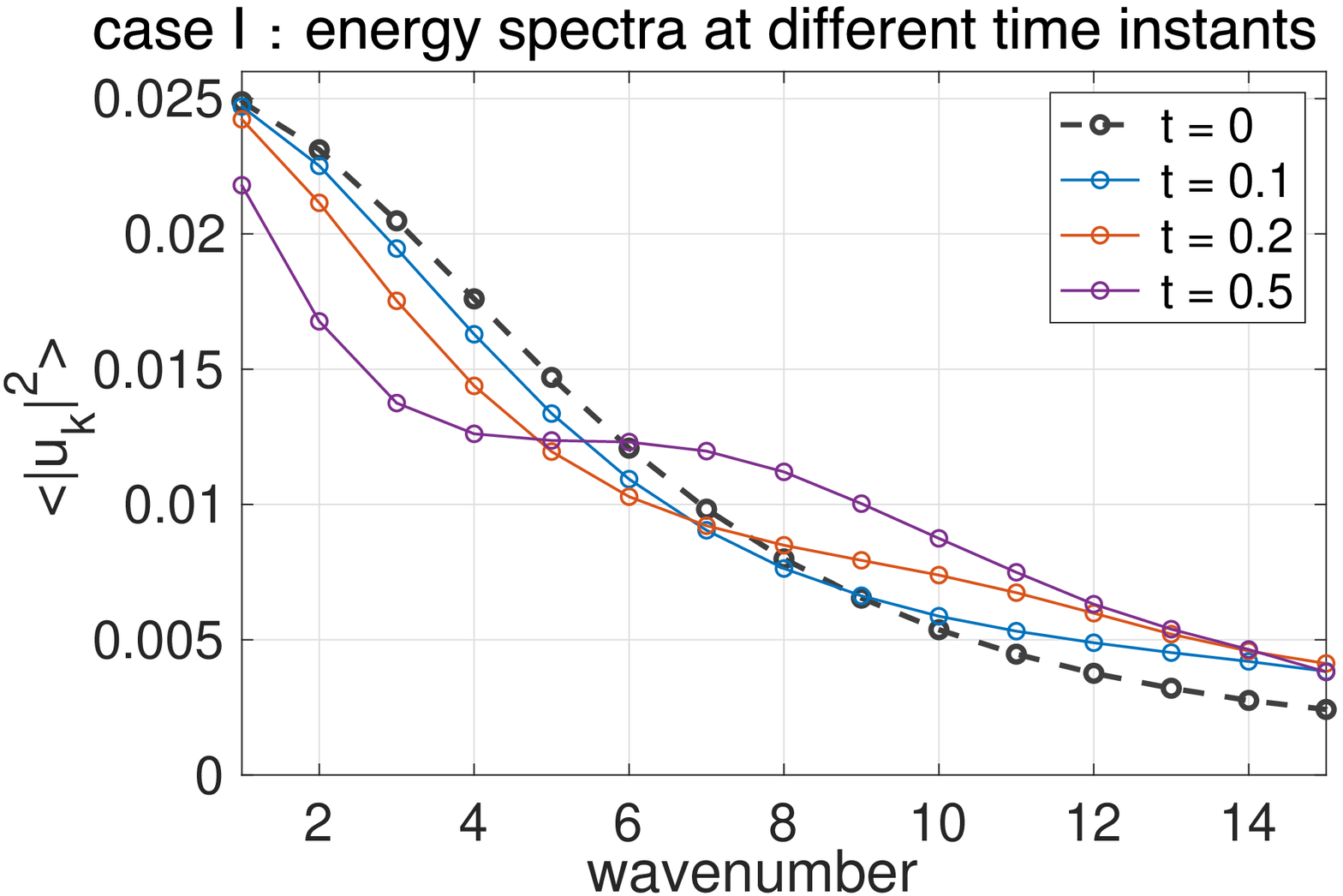}\includegraphics[scale=0.29]{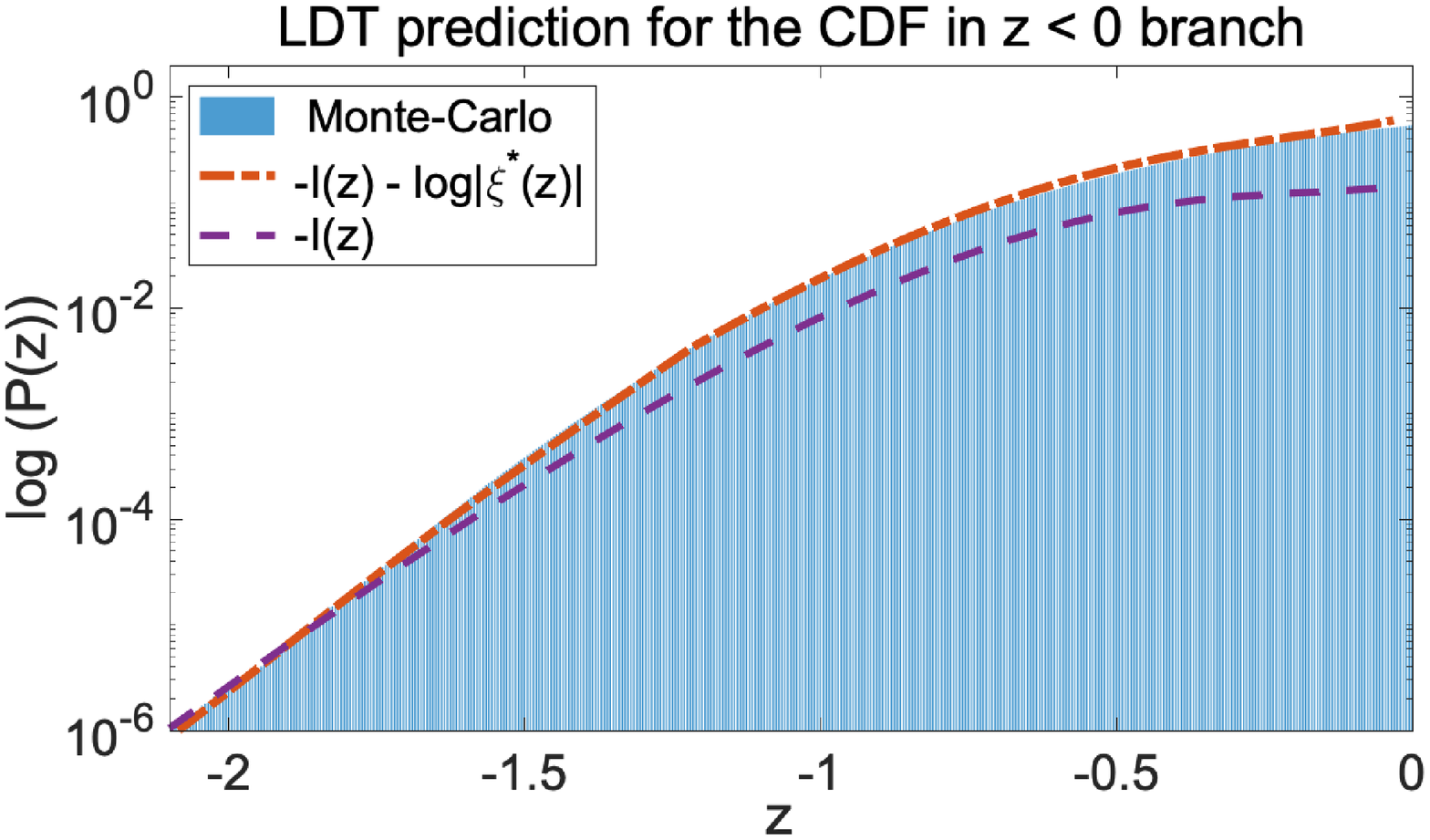}\includegraphics[scale=0.29]{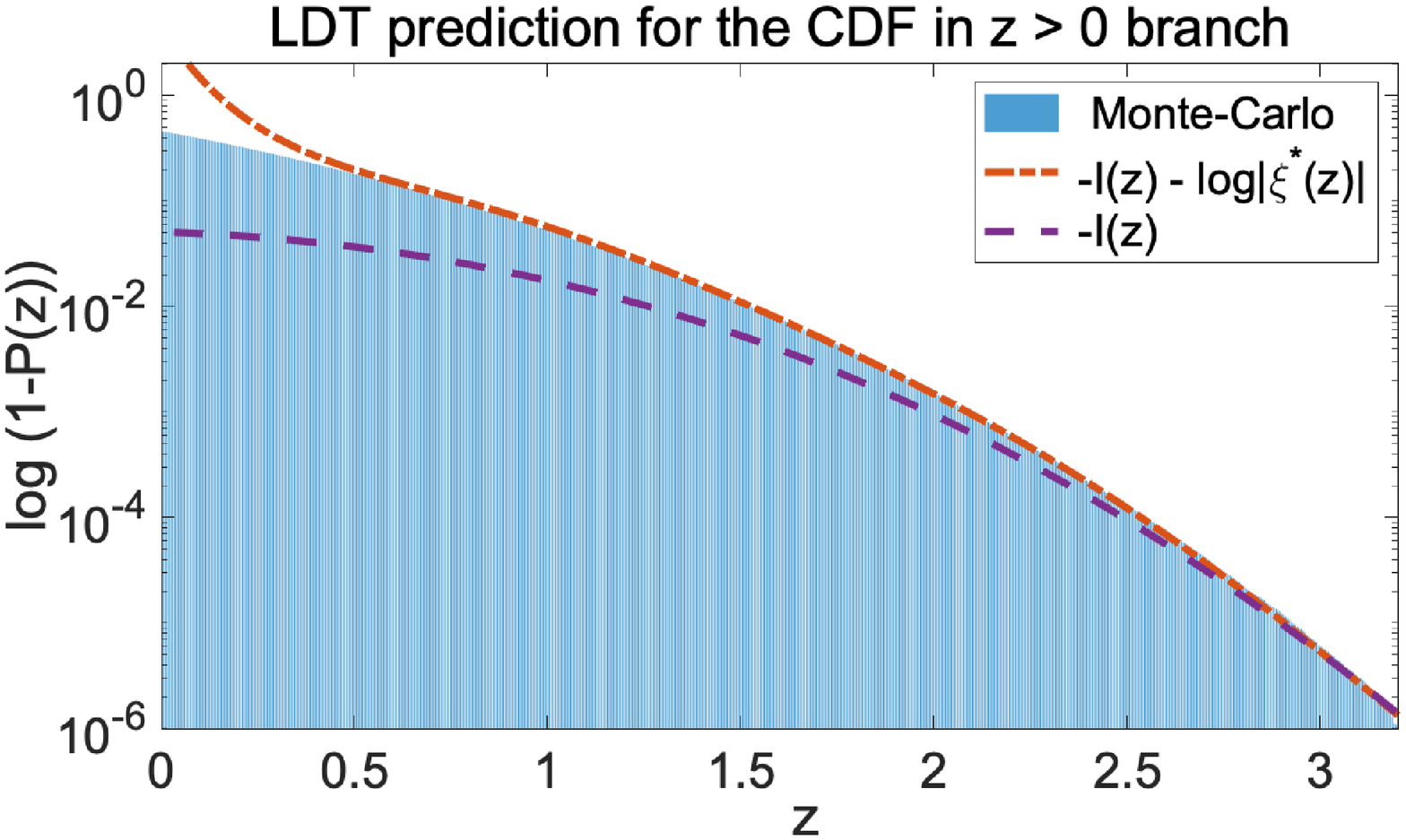}}

\subfloat[Case II: equipartition initial energy spectrum]{\includegraphics[scale=0.29]{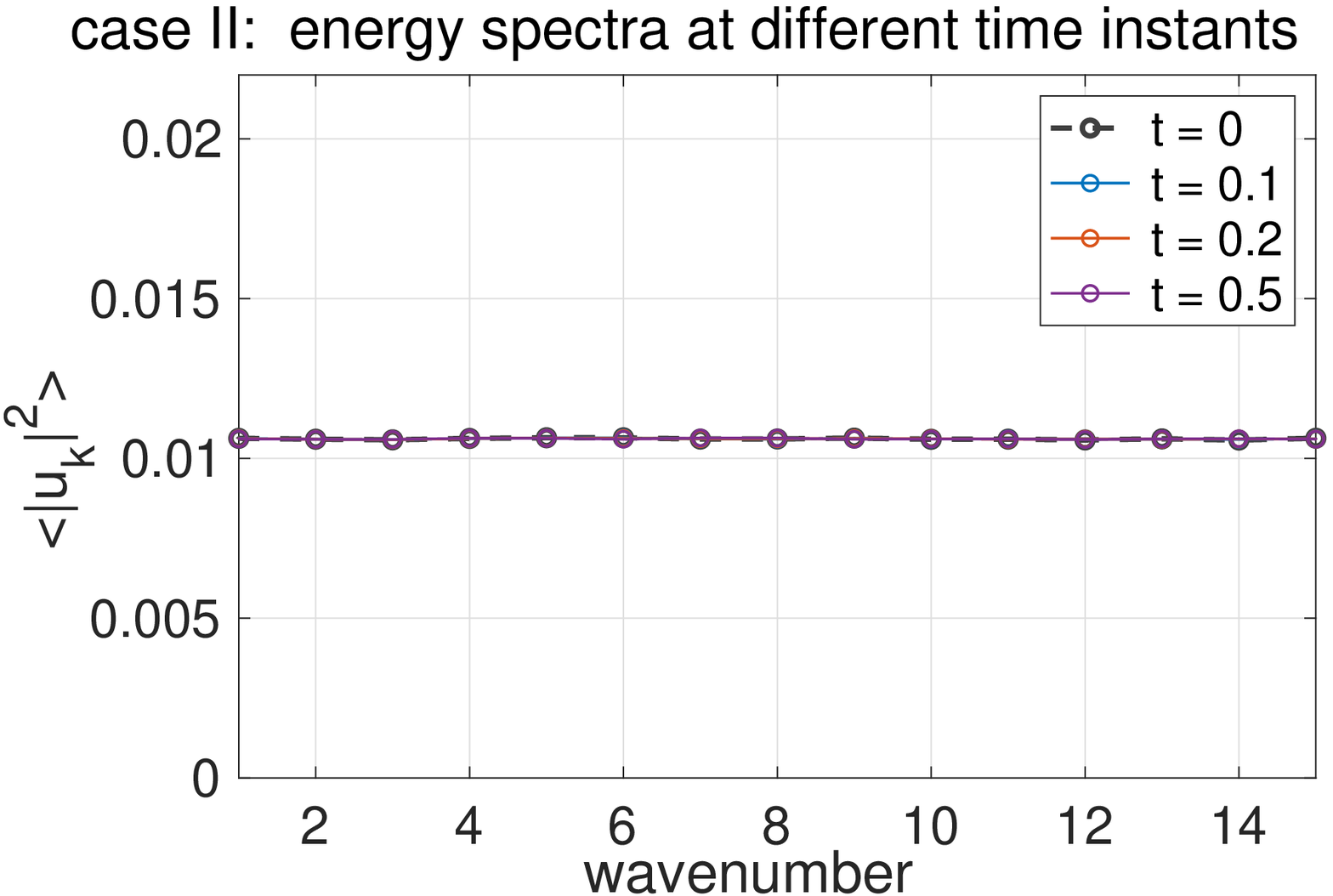}\includegraphics[scale=0.29]{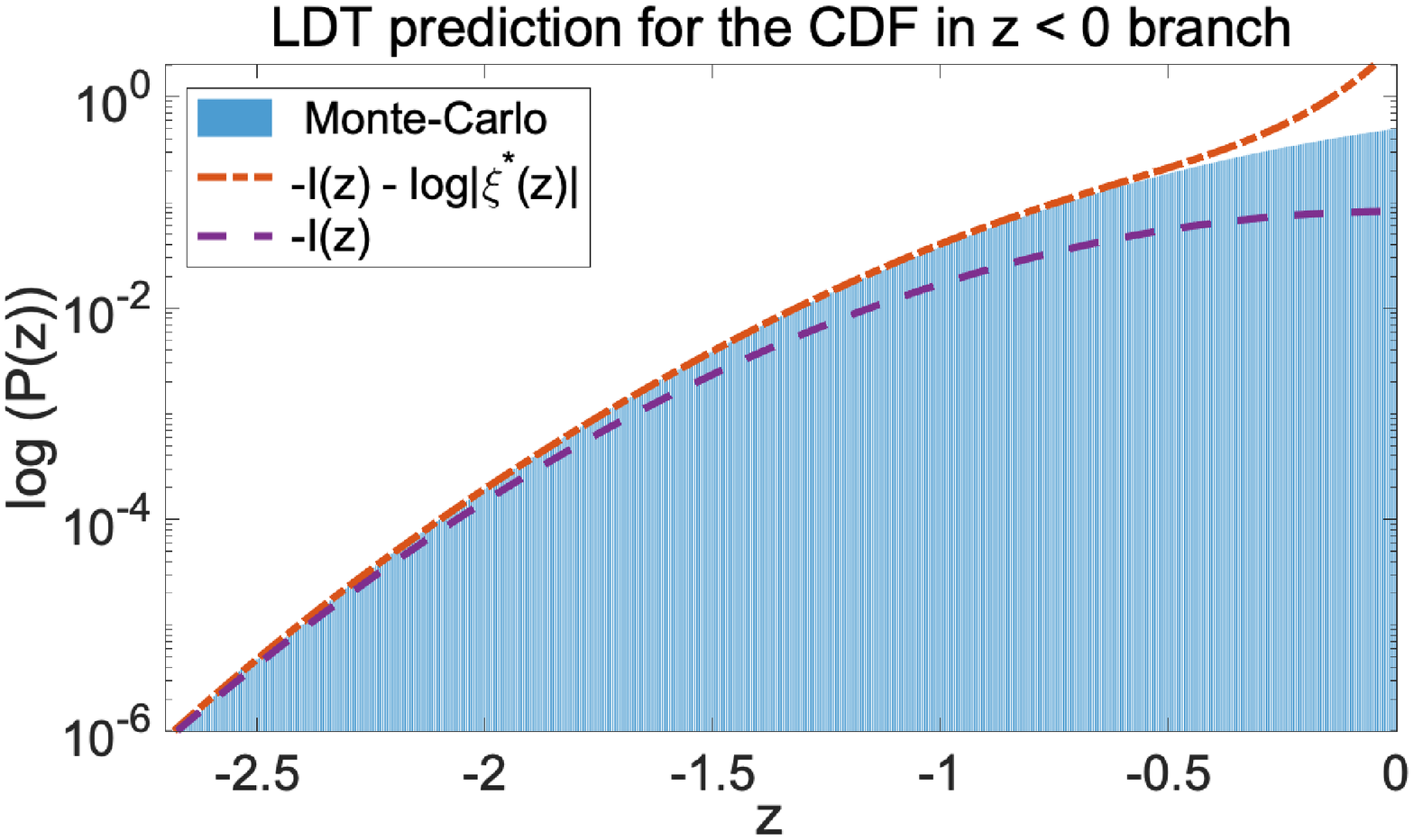}\includegraphics[scale=0.29]{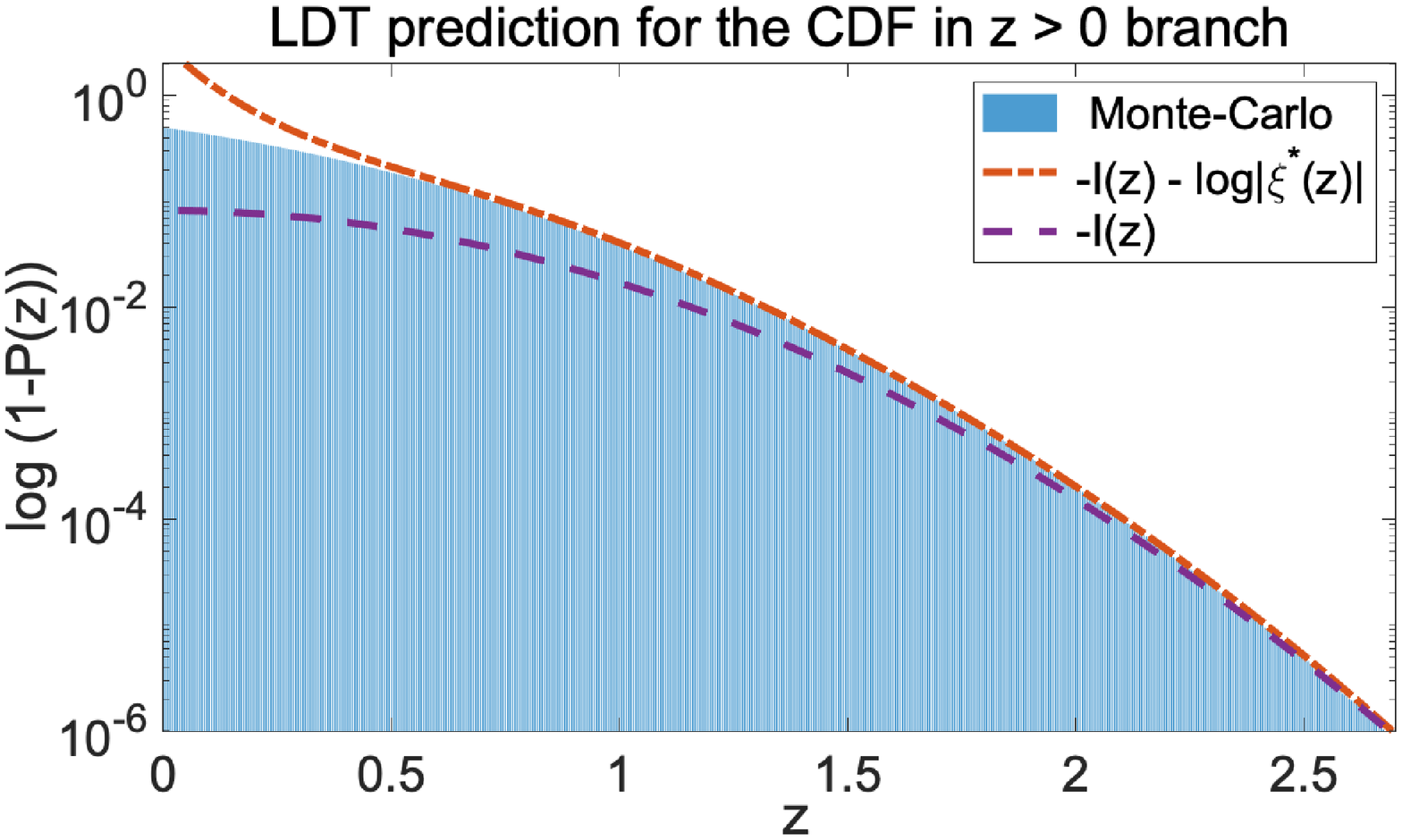}}

\caption{Energy spectra of the TKdV solutions
$u_{\Lambda}\left(t,x_{c}\right)$ from direct Monte-Carlo simulations, and LDT predictions of the probability distributions in
the negative branch $\mathbb{P}\left\{ u\protect\leq-z\right\}$
(left) and the positive branch $\mathbb{P}\left\{ u\protect\geq z\right\}$
(right) in the two test regimes with different initial data. The direct
Monte-Carlo results as the target are shown by the histogram. The
leading order LDT prediction $-I_{\Lambda}\left(z\right)$ is compared
with the additional second-order correction $-\log\left|\xi^{*}\right|$.\label{fig:LDT-predictions-CDF}}
\end{figure}

The predictions of the probabilities in the positive branch $\mathbb{P}\left(u\left(T\right)\geq z\right)$ and the negative branch $\mathbb{P}\left(u\left(T\right)\leq-z\right)$ at the final equilibrium time $T=1$ are shown in Figure \ref{fig:LDT-predictions-CDF} for a wide range of the extreme values $z$. The leading order approximation from $\exp\left[-I_{\Lambda}\left(z\right)\right]$ in (\ref{eq:LDT}) is compared with the refined estimate in~(\ref{eq:expansion_gau}), showing that the latter accurately matches the results from Monte-Carlo simulations for a wide range  of  positive and negative values of $z$.
Comparing the results in the two cases of initial distribution, we see that case I, with a decaying energy spectrum,  generates slower decay rate in the positive side $z>0$, and the prediction range is narrower with faster decay in the negative branch $z<0$. In case II, which starts from equipartition of energy, the two branches decay at similar rate with comparable probability to either positive or negative extreme values. This leads to the development of a skewed asymmetric PDF in case I, while a relatively symmetric PDF is reached in case II.

\subsubsection{LDT prediction of PDFs at different time instants }

\begin{figure}
\subfloat[Case I: decaying initial energy spectrum]{\includegraphics[scale=0.33]{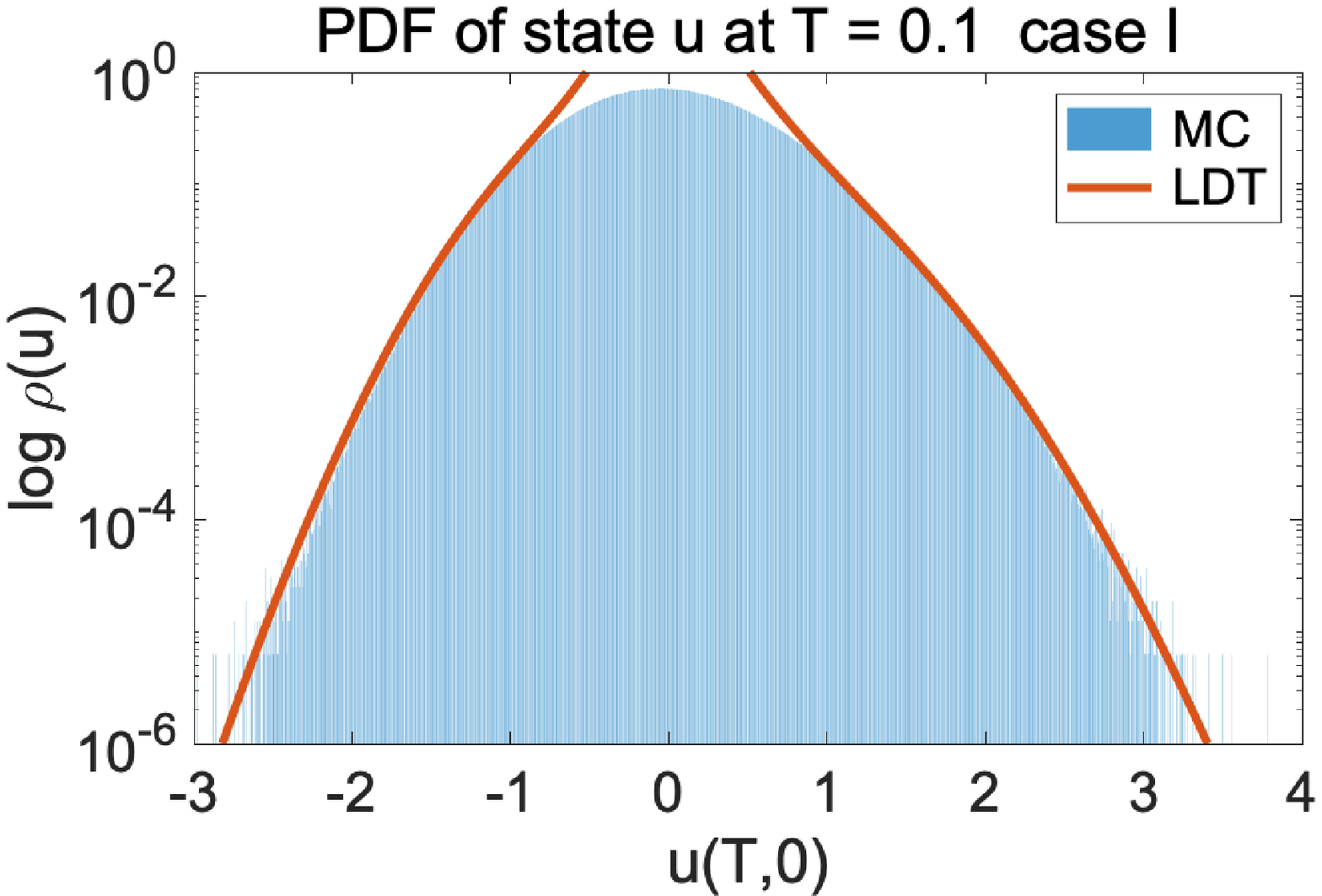}\includegraphics[scale=0.33]{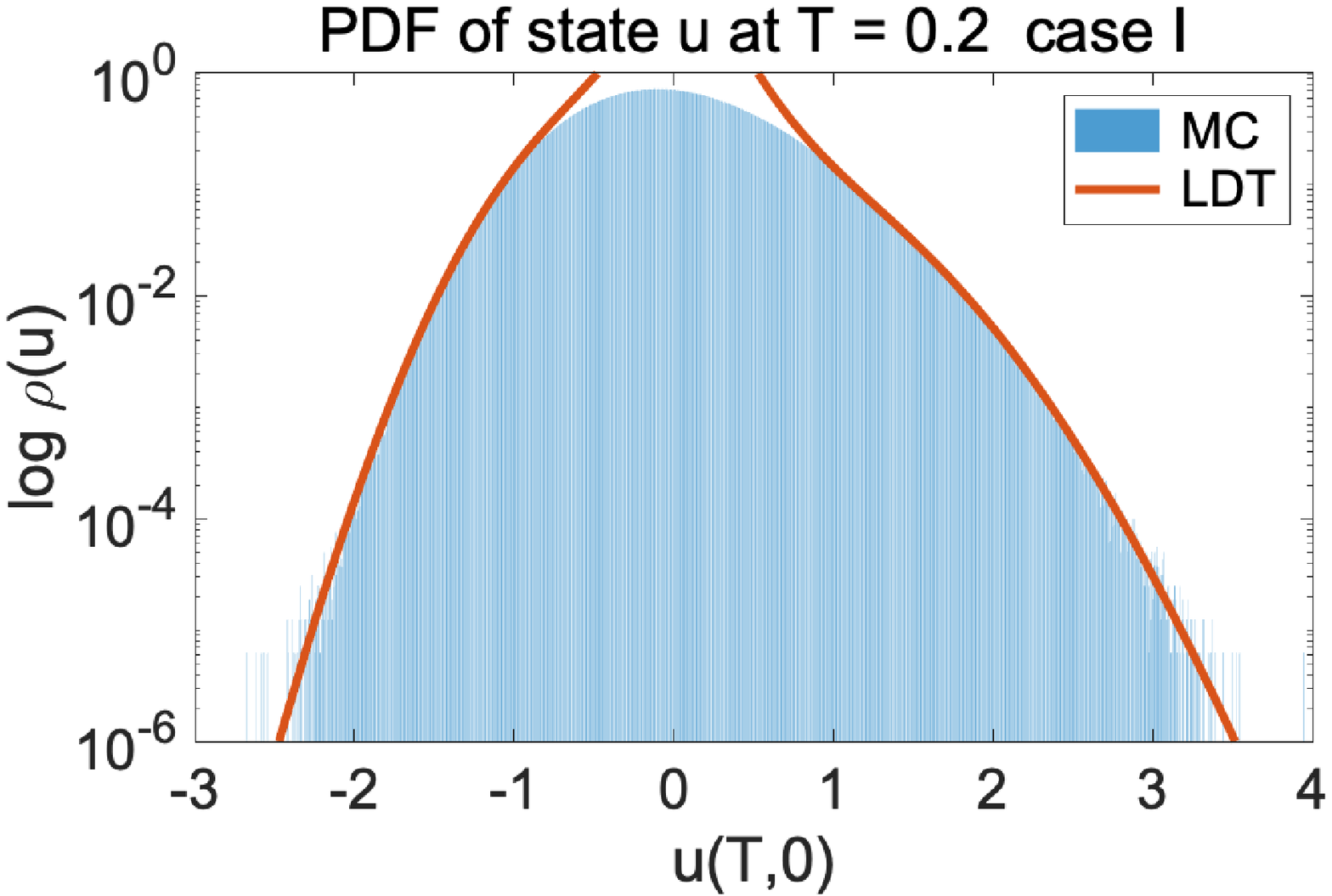}\includegraphics[scale=0.33]{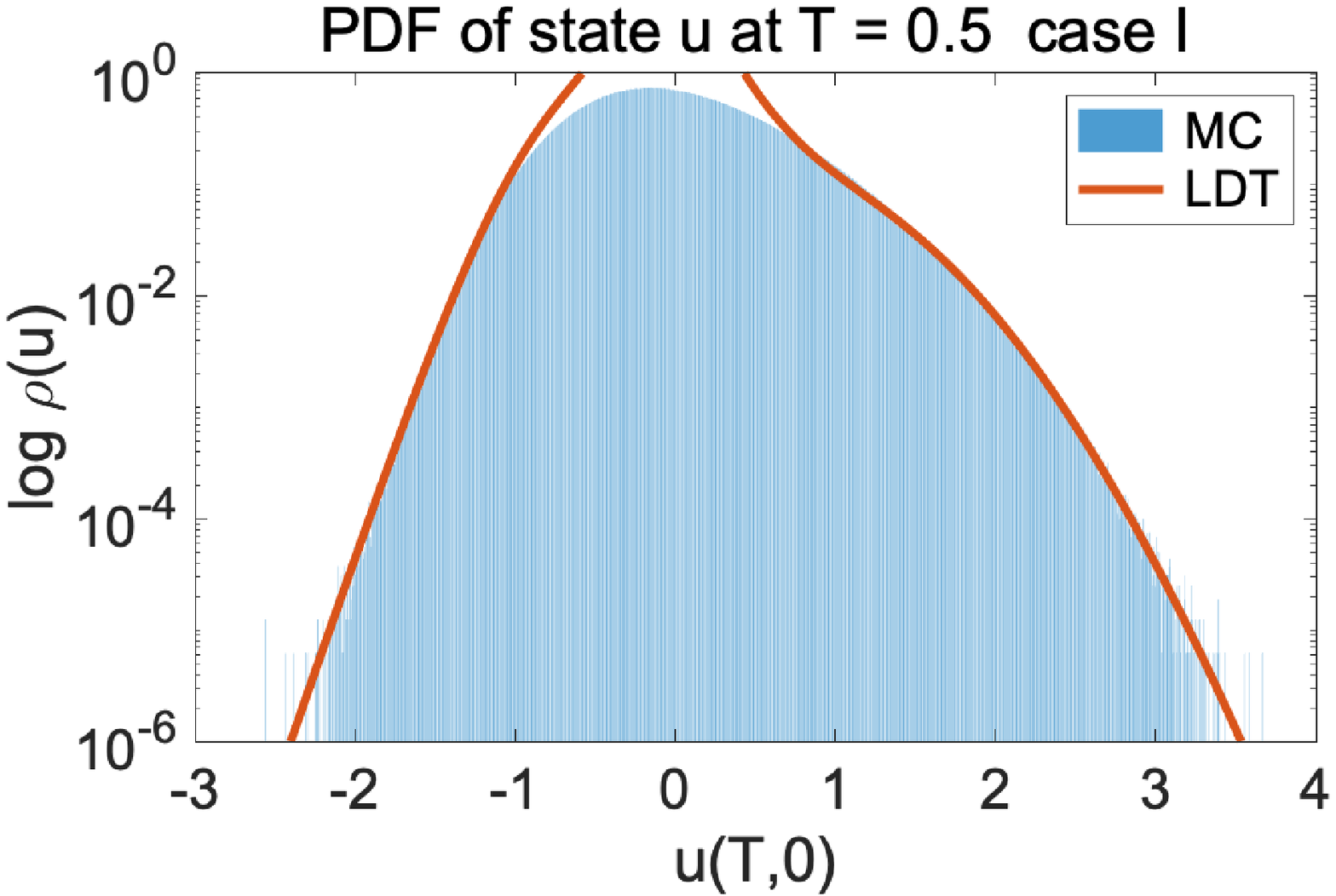}
}

\subfloat[Case II: equipartition initial energy spectrum]{\includegraphics[scale=0.33]{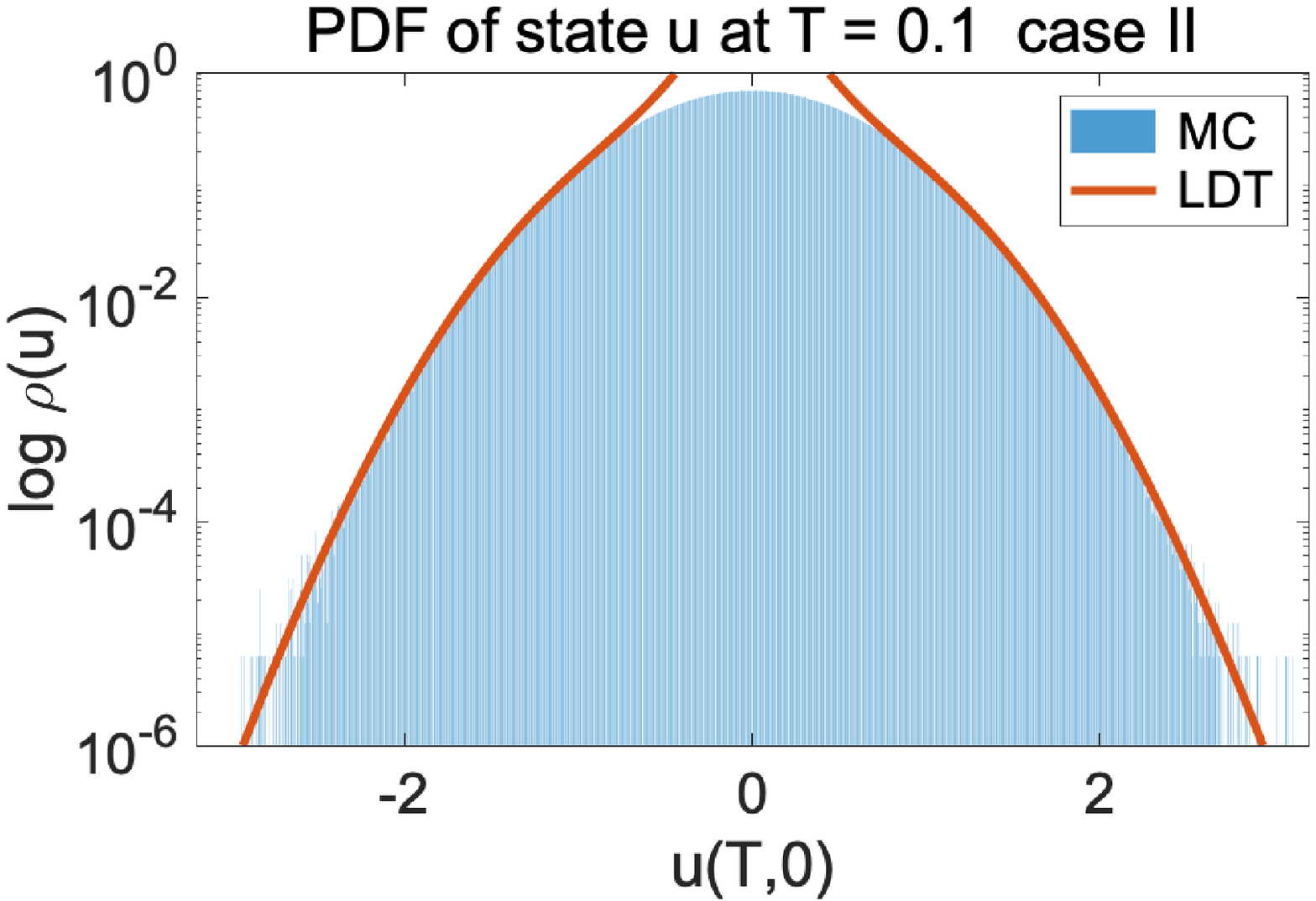}\includegraphics[scale=0.33]{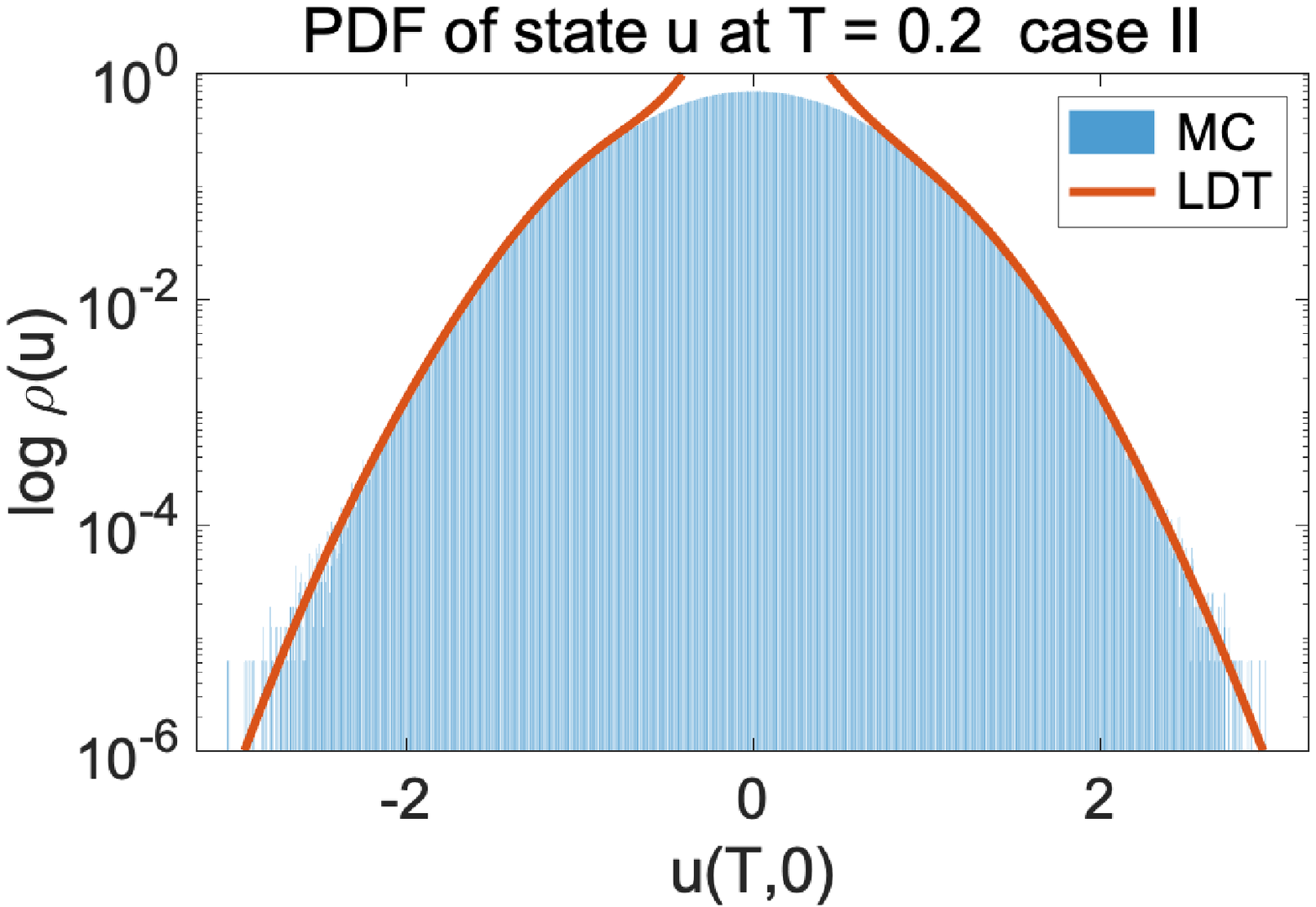}\includegraphics[scale=0.33]{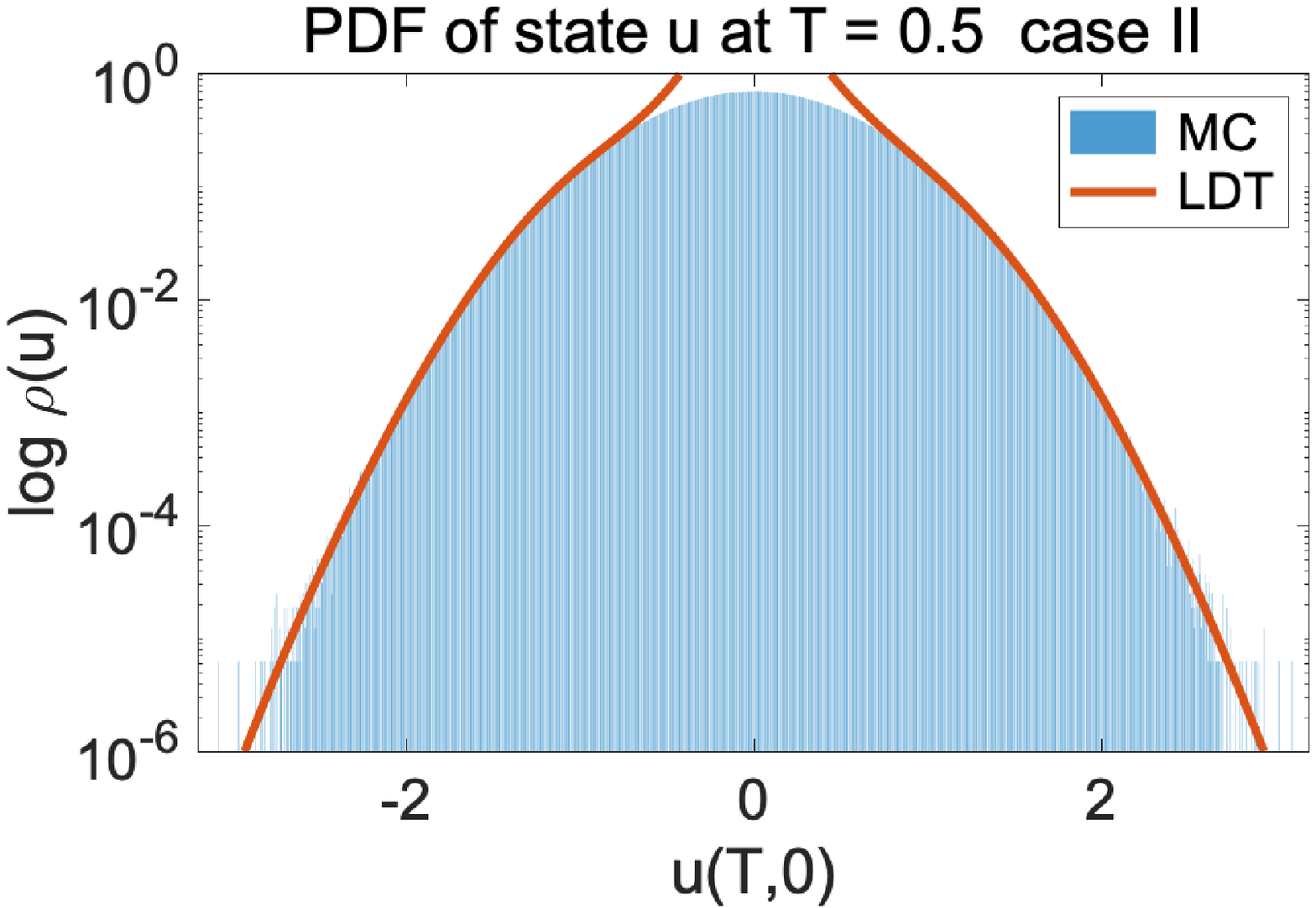}
}

\caption{Comparison of the PDFs of the TKdV solution $u_{\Lambda}\left(t,x_{c}\right)$
from direct Monte-Carlo simulation and the large deviation prediction
at different time instants $t=0.1,0.2,0.5$ before the final equilibrium.
The two test cases in Gaussian initial data with a decaying energy
spectrum and equipartition energy spectrum are compared. \label{fig:Comparison-of-PDFs}}
\end{figure}

Next, we can use the LDT result for probability distributions to discover
the time evolution of the PDFs during the starting
transient state before the equilibrium state is reached. The explicit formula to compute the PDFs is shown in (\ref{eq:pdf}) of the Appendix. The PDFs
of the TKdV state $u_{\Lambda}\left(t,x_{c}\right)$ at several different
time instants $t=0.1,0.2,0.5$ are compared in Figure \ref{fig:Comparison-of-PDFs}.
In the LDT predictions, the additional refined corrections~\eqref{eq:expansion_gau}
are needed to accurately capture the shapes of the distribution functions.
Different shapes for the positive and negative side of PDF
tails with distinct statistics are developed in time based on the
two initial energy spectra. The equipartition energy case is simpler
with near-symmetric tails on positive and negative sides.
The decaying energy spectrum case is more challenging with the gradual
development of a skewed PDF. LDT prediction is able to capture
the development of PDF tails in both cases at different time instants. Even
starting from not very large values of $z$, the long tails of
the PDFs are estimated with high accuracy by the LDT predictions in both
positive and negative sides of the values. In particular, the LDT can
keep the precision for a much wider range of extreme values without much increase of computational cost, while
the Monte-Carlo simulation will become too expensive to resolve the long 
tails in the PDFs with desirable accuracy.

\subsection{Trajectories to extreme events with different amplitudes}
\begin{figure}

\subfloat[trajectory to negative extreme events]{\includegraphics[scale=0.4]{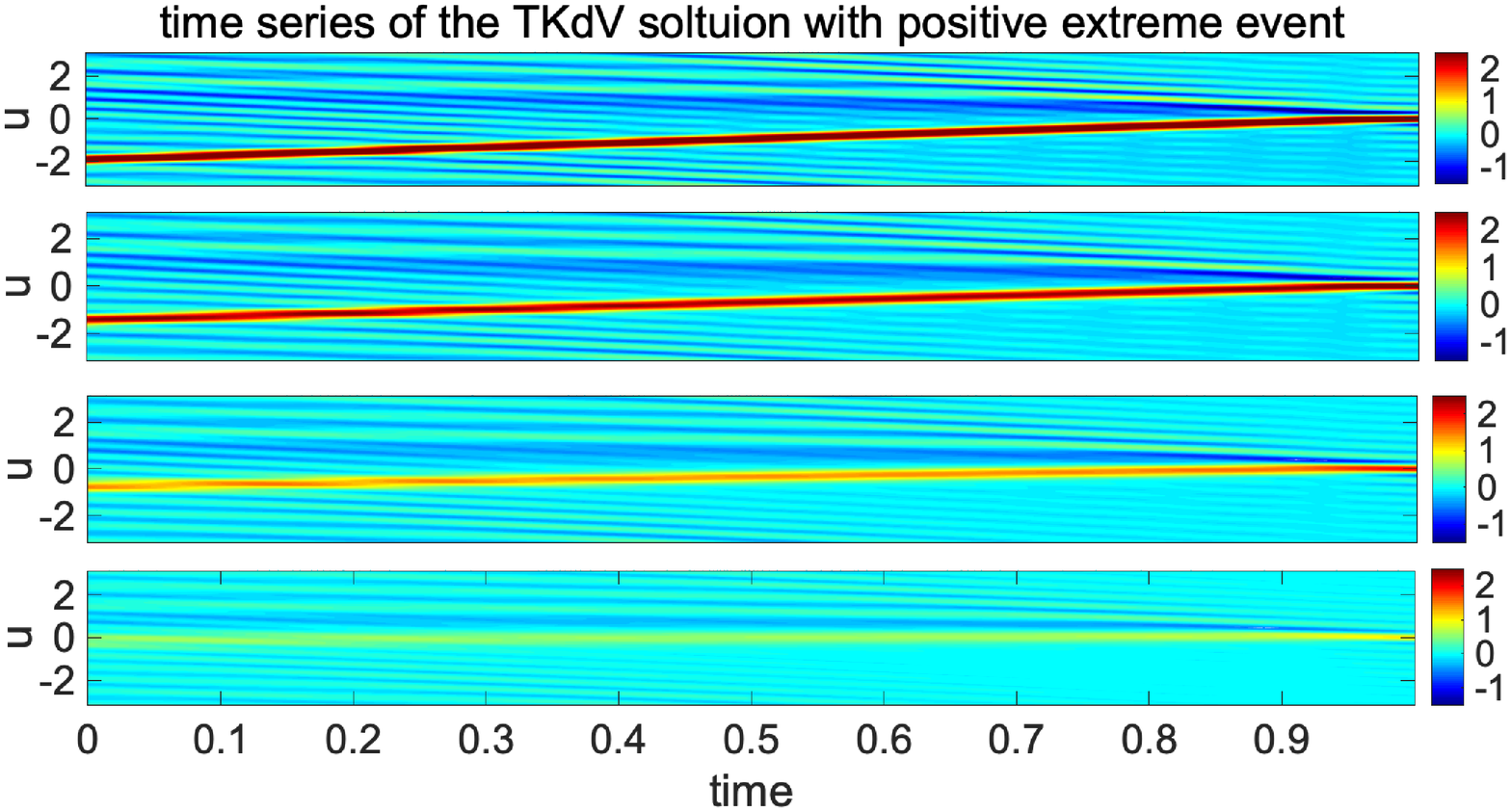}
\includegraphics[scale=0.4]{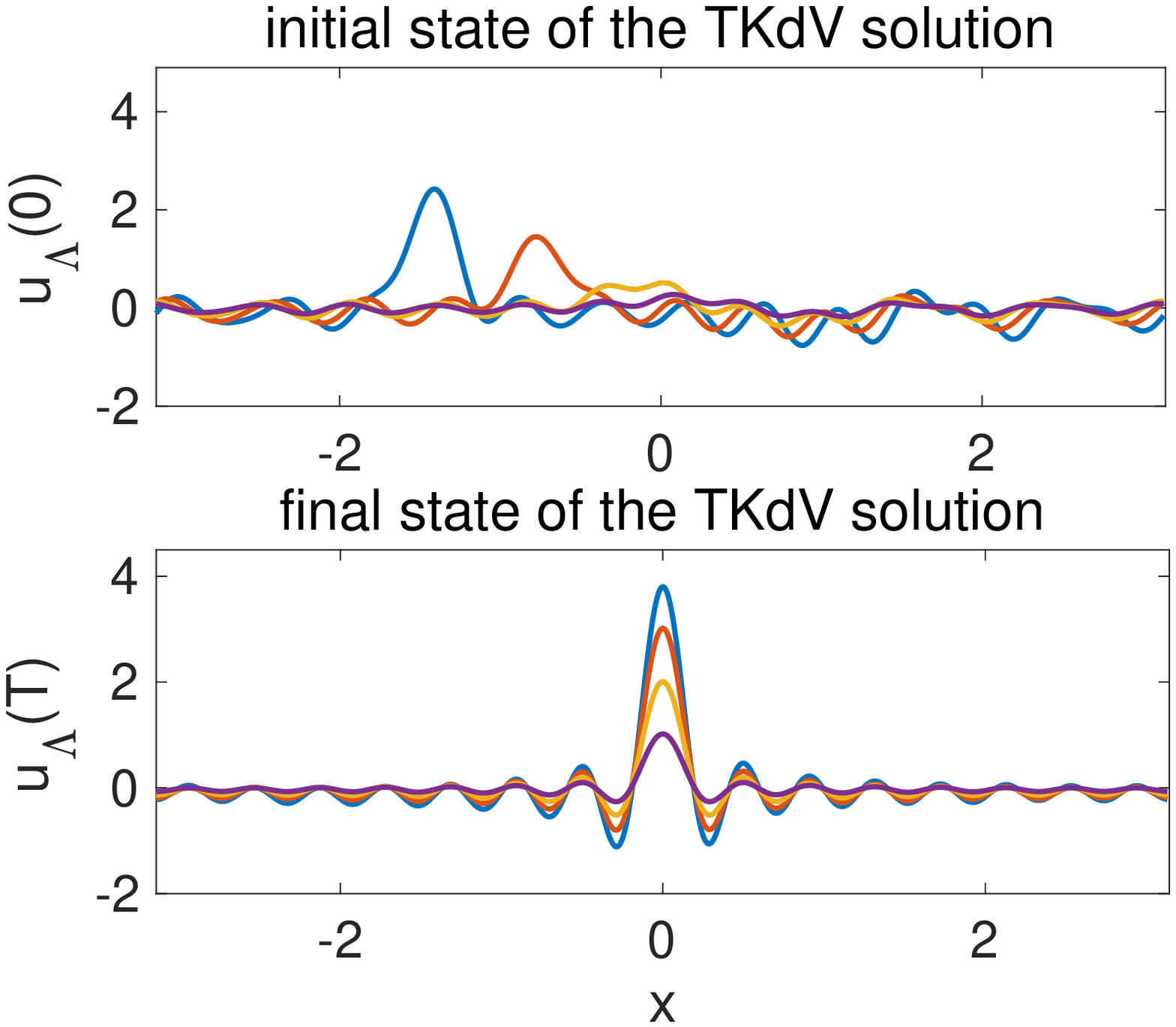}}

\subfloat[trajectory to negative extreme events]{\includegraphics[scale=0.4]{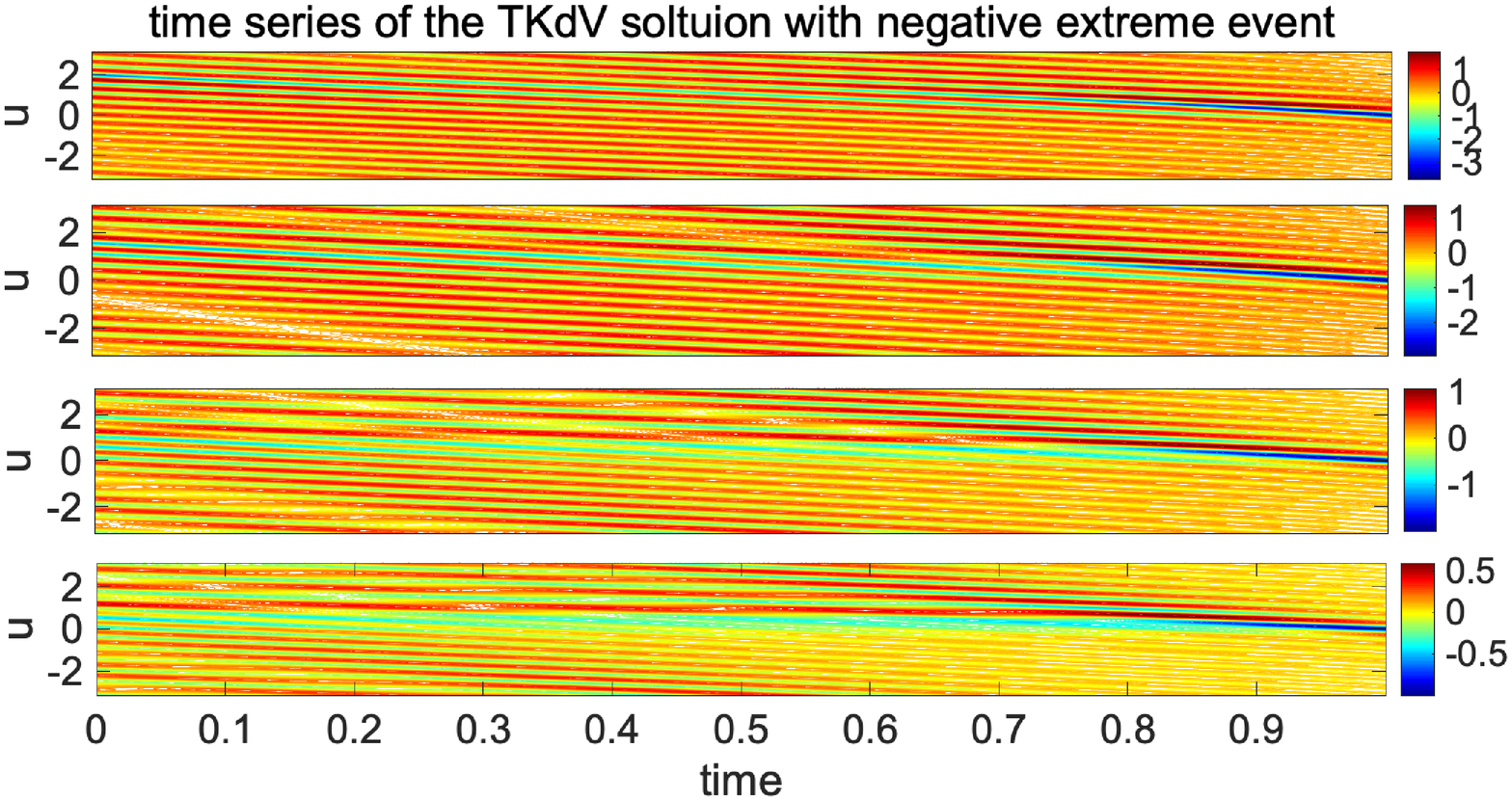}
\includegraphics[scale=0.4]{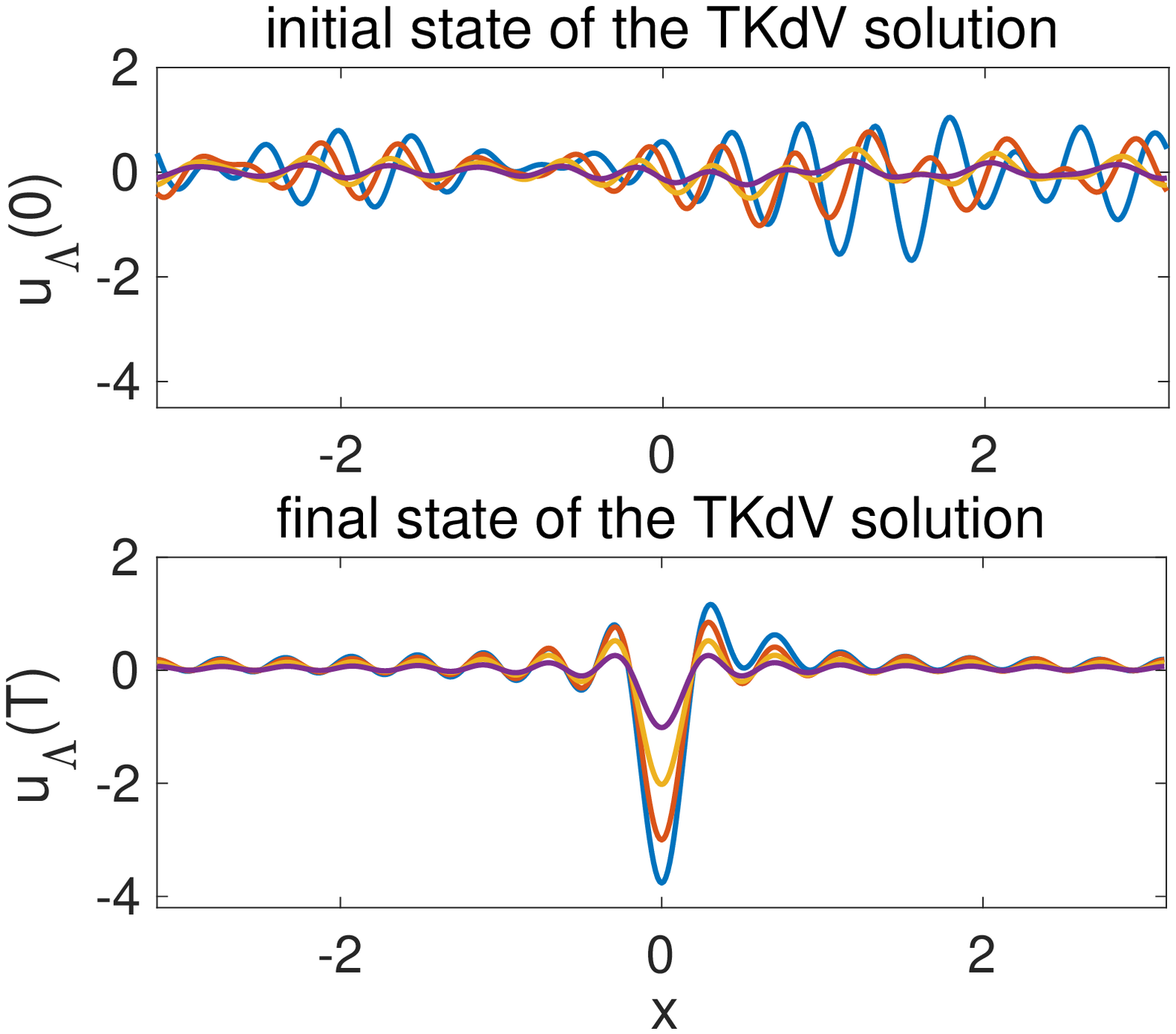}}

\caption{Time evolution of the LDT trajectory of the TKdV solution starting from random Gaussian initial data to positive and negative extreme events. The initial and final solution profiles are compared in the right panel. Different dynamical structures are observed in the route to positive and negative extreme values.\label{fig:Optimized-trajectory}}
\end{figure}

The LDT prediction estimates the probabilities of extreme events not only in the final equilibrium state but also during the intermediate transient states. Next, we focus on case I with skewed PDF and check the characteristic trajectories that lead to extreme values on the positive and negative sides. 
In Figure \ref{fig:Optimized-trajectory} we plot the LDT critical solutions with both positive and negative extreme events to reach different values $z$ at the final time $T=1$. On the left panel, the LDT trajectories $u_{\Lambda}\left(t,x\right)$ to reach final extreme values $z=\pm4,\pm3,\pm2,\pm1$ are compared. It is interesting to contrast the routes to positive and negative extreme events in the typical trajectories with quite different dynamical evolution. To reach an increasingly large positive value, a dominant localized wave package emerges in the initial state. The wave package travels at different wave speeds and interacts with many fluctuating small-scale structures to converge to the extreme value in the center of the domain at the final time. On the other hand, to form a large negative value, multiple small-scale interacting wave patterns with high wavenumbers are created to enforce the negative peak to emerge at the very last stage. 

On the right panel of Figure \ref{fig:Optimized-trajectory}, we show the snapshots of the initial and final state of the solution $u_{\Lambda}$ with different final extreme values. Consistent with the previous observation, reaching a large value of value $z$ at time $T=1$ requires the formation of a relatively strong localized jet in the initial state $u_{\Lambda}\left(0,x\right)$ for the positive extreme event; in contrast oscillatory initial waves are required to drive to a dominant negative extreme event. From the four typical LDT trajectories, we can observe how an extreme large events are gradually accumulated from a concatenate series of events. These characterizing structures in LDT solutions agree with the typical trajectories with positive and negative extreme events shown in Figure \ref{fig:Solutions-of-tKdV} from direct simulations.

\subsubsection*{Asymmetry in the extreme values on the positive and negative side}

\begin{figure}
\begin{centering}
\includegraphics[scale=0.4]{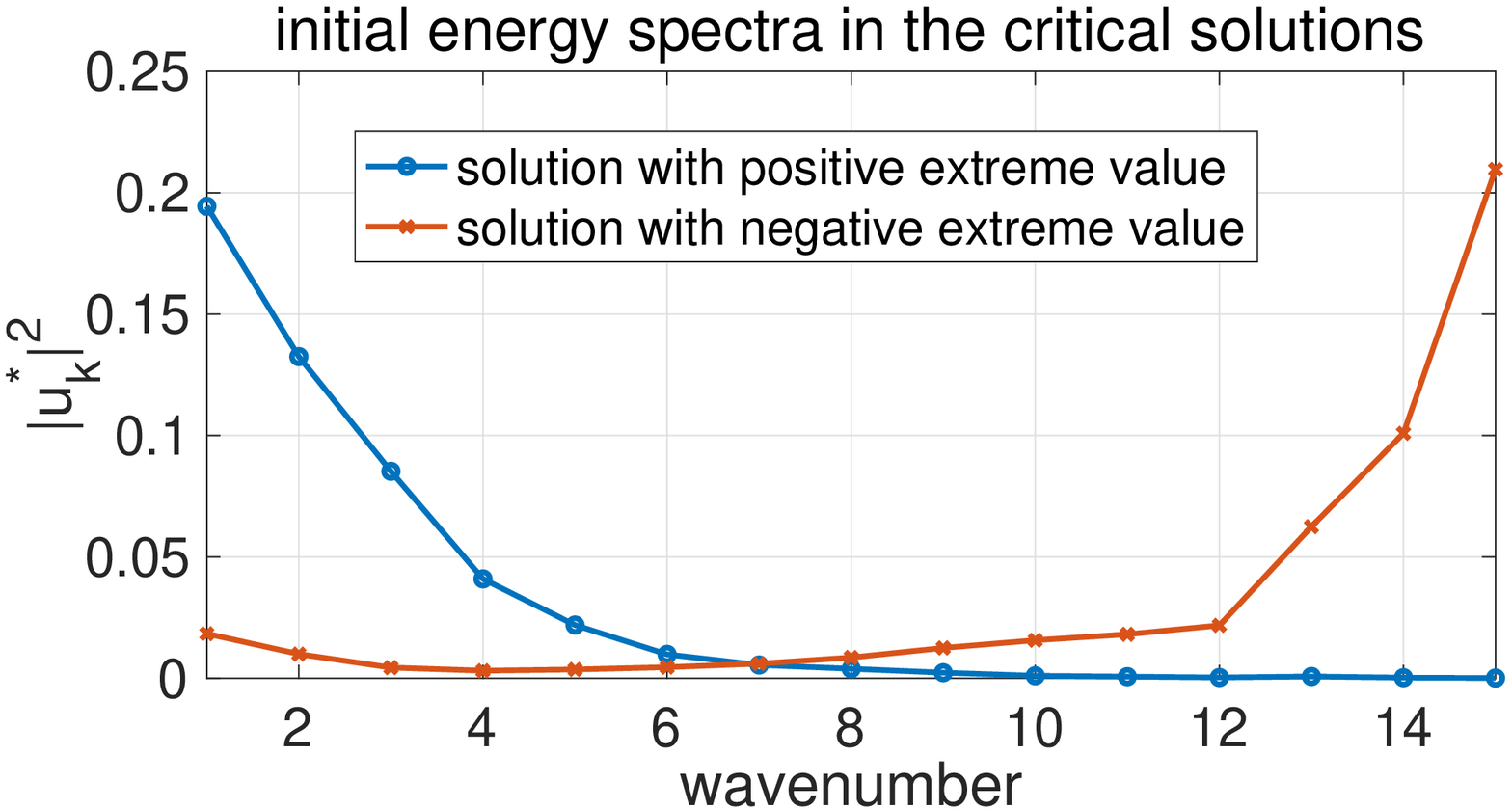}\includegraphics[scale=0.4]{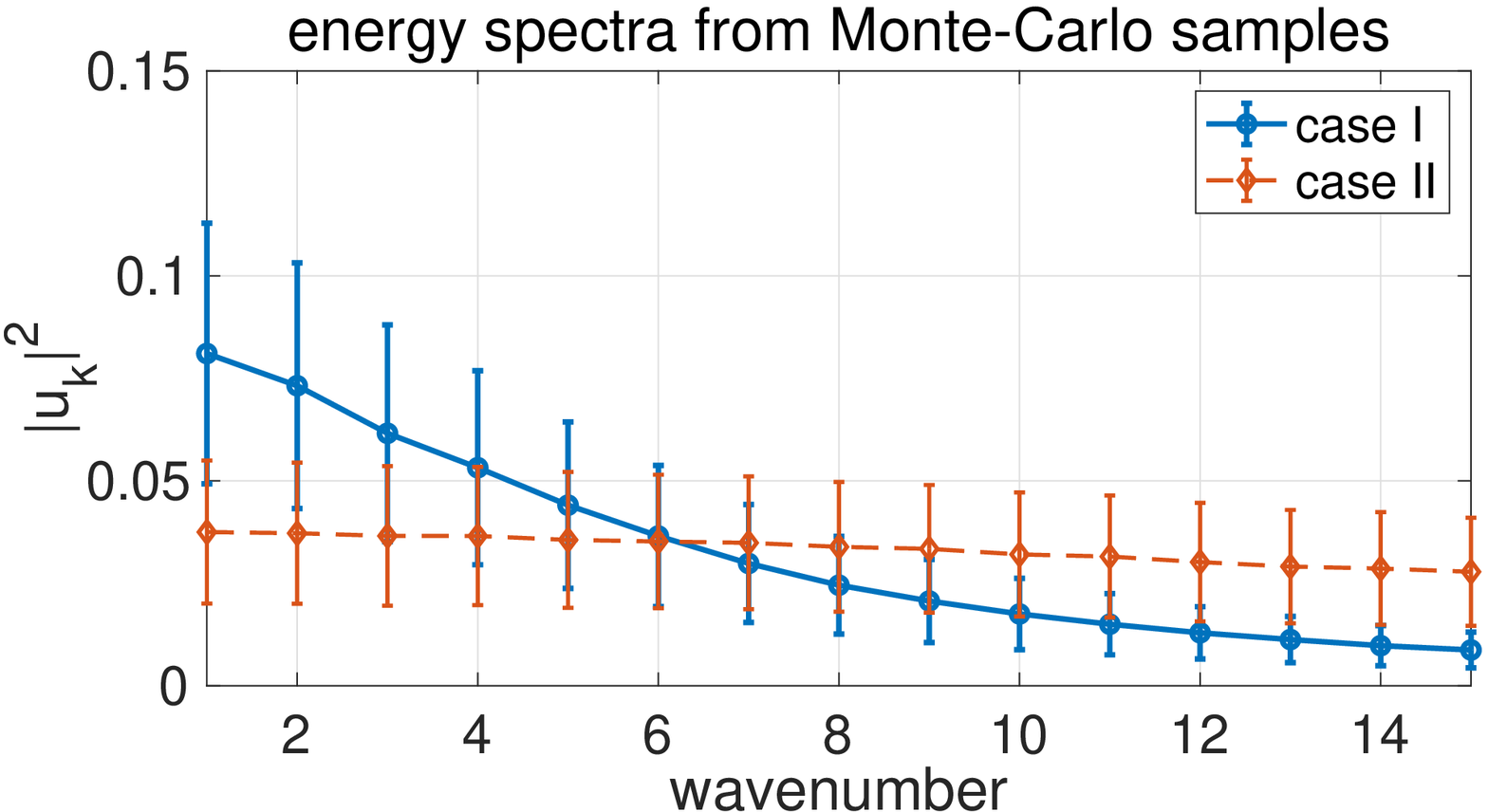}
\par\end{centering}
\caption{Left: normalized energy spectra of the LDT solutions to reach positive and negative extreme values; Right: the energy
spectra computed from samples of the direct Monte-Carlo simulation.
The standard deviation from the energy in the mode among the samples
is shown in the error bar.\label{fig:spec_comp} }
\end{figure}

The LDT solutions for positive and negative extreme events in Figure \ref{fig:Optimized-trajectory} provide useful insights for the asymmetric PDF structure developed in the final state. To see this, on the left panel of Figure \ref{fig:spec_comp} we show the initial energy spectra of the LDT solutions with positive and negative extreme events. It illustrates the required energy variability level in each initial mode to develop extreme event at $T=1$. 
For a positive extreme value to form up, a decaying energy spectrum is required in the initial time; in contrast an increasing energy spectrum with more energetic small-scale modes is needed for a negative extreme event to develop.  The shape of the initial energy spectrum in the LDT solution provides necessary condition for a positive or negative extreme event to develop in the future time. 

In the right panel of Figure \ref{fig:spec_comp} we compare the statistical energy spectra from the MC samples from different energy spectra.
We see that the decaying energy spectrum (case I) gives a larger variability among the large scales. Therefore, a positive extreme event is easier to reach from this initial state while the negative extreme event requires large energy in small scale modes that is difficult to occur. A positively skewed PDF is then expected in the final equilibrium. On the other hand, with an equipartition of energy (case II), the initial spectrum is flat with similar amplitudes in both small and large scales. As a result, the required configurations for positive and negative extreme events become equally likely to be reached, thus leading to a symmetric final distribution.

\section{Numerical tests for the LDT prediction with Gibbs initial data\label{sec:Numerical-tests-Gibbs}}

Here, we use the LDT approach  to compute the statistics of solution emerging from the non-Gaussian Gibbs distribution~\eqref{eq:gibbs_invar}. We pick the value $\theta=1$ in the test, which leads to solutions whose distribution develops a strong skewness during the evolution.  
The numerical setup of the TKdV model (\ref{eq:model}) is kept the same as the previous Gaussian case in Sect.~\ref{sec:Numerics-to-compute}.

\subsection{Converged LDT solutions with different values of $\mu$ and $\lambda$}

\begin{figure}
\centering
\includegraphics[scale=0.32]{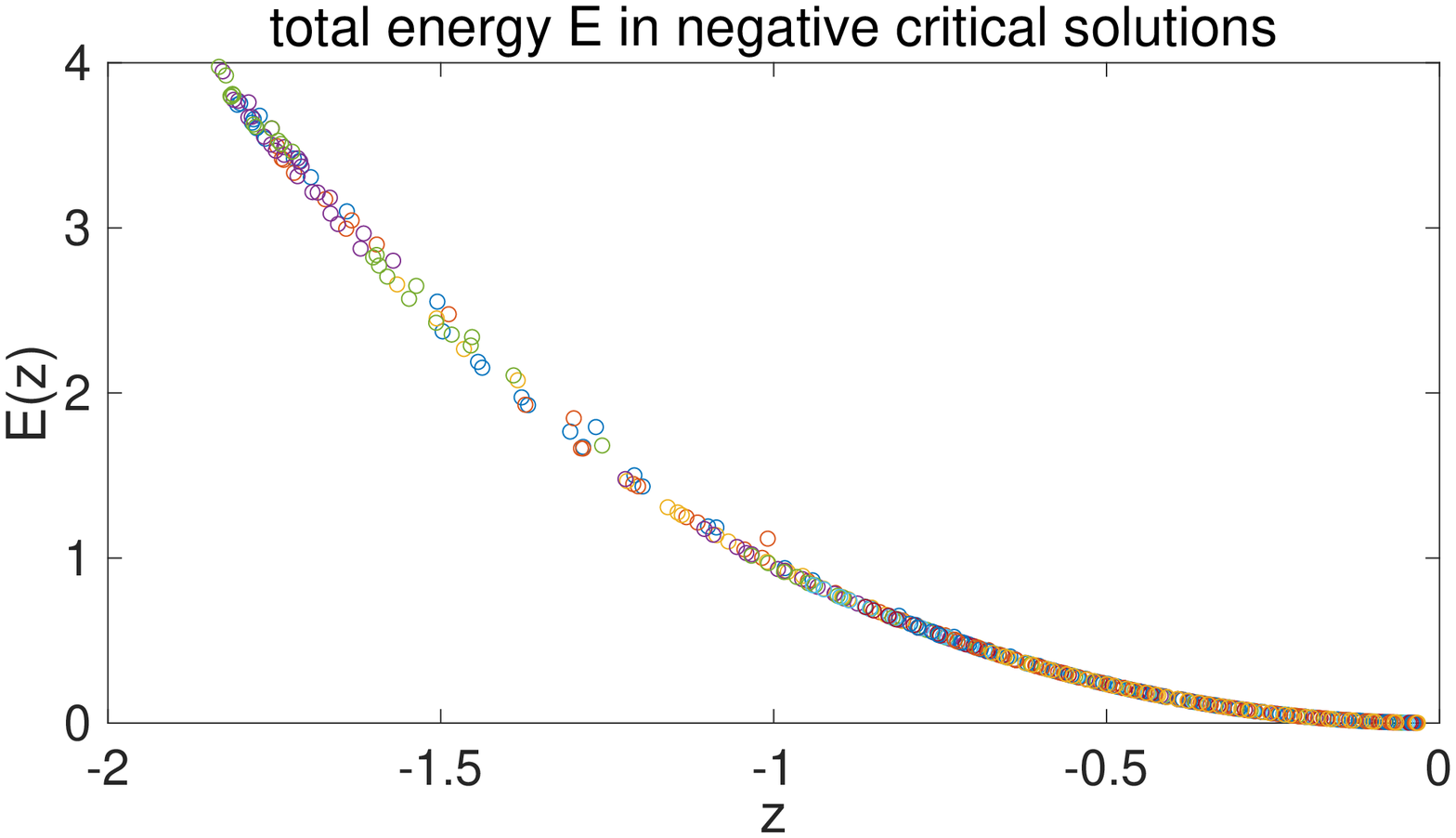}\includegraphics[scale=0.32]{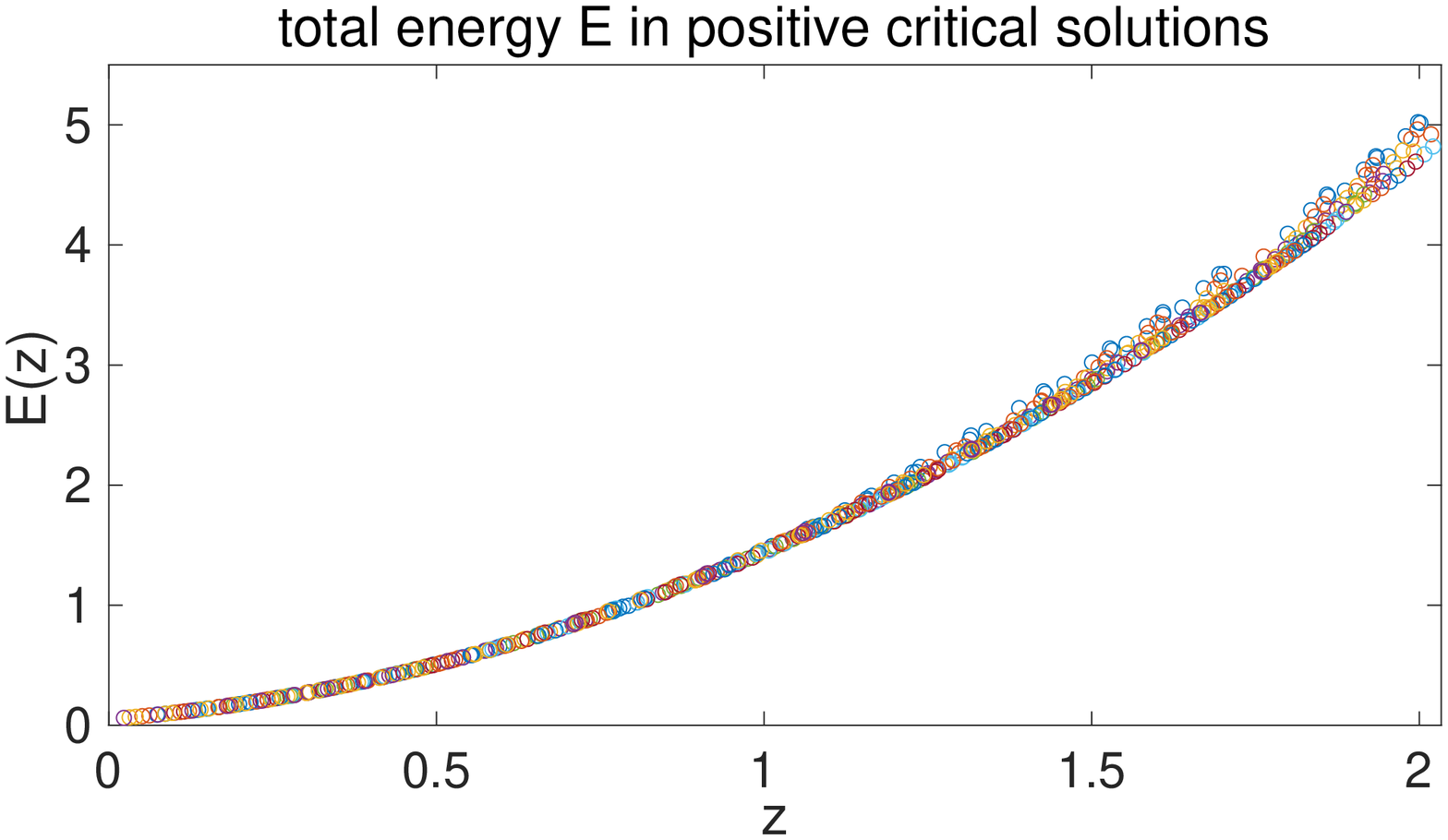}

\caption{The maximum energy $E\left(z\right)=\frac{1}{2}\left\Vert u^{*}_{\Lambda}\right\Vert ^{2} $
in the LDT solutions along the optimization process with different
extreme values of $z$ using different constant values of $\mu$.
Markers with different colors represent the converged LDT solutions
from different values of $\mu$.\label{fig:The-total-energy}}
\end{figure}

First, we need to discuss the choice of the two Lagrangian multipliers $\mu$ and $\lambda$ in the Gibbs case optimization problem (\ref{eq:optim_gibbs}).
The new additional parameter $\mu\left(z\right)$ is picked to control the saturated value of total energy $E$ based on the extreme value of $z$. From the formulas in Sec.~\ref{subsec:gibbs}, we
find that $\mu$ should be increased to larger values as $\lambda$
(thus $z$) grows large in order to reach a converged LDT solution
for finite $z$ in the optimization scheme. In the practical numerical
computation, for convenience of choice, we empirically pick a constant
value of $\mu$ during the optimization process for different extreme
values of $z$ (that is, to use different values of $\lambda$). This
is valid by the assumption that $\mu\left(z\right)\simeq\mu$ is not very sensitive around
some range of values of $z$, thus the converged solution is not much
affected by small changes in the value of $\mu$. To confirm this, 
we compare the total energy $E\left(z\right)=\frac{1}{2}\left\Vert u^{*}_{\Lambda}\right\Vert ^{2} $
in the converged LDT solutions $z=u^{*}_{\Lambda}\left(T;\lambda,\mu\right)$
achieved from different values of $\mu$ in Figure \ref{fig:The-total-energy}. For each fixed value of
$\mu$, the extreme event value is computed by changing the value
of $\lambda$. It can be seen that the solution converges to the total
energy of similar range using different values of $\mu$. 

\subsection{Prediction of skewed CDFs and PDFs}

\begin{figure}
\includegraphics[scale=0.29]{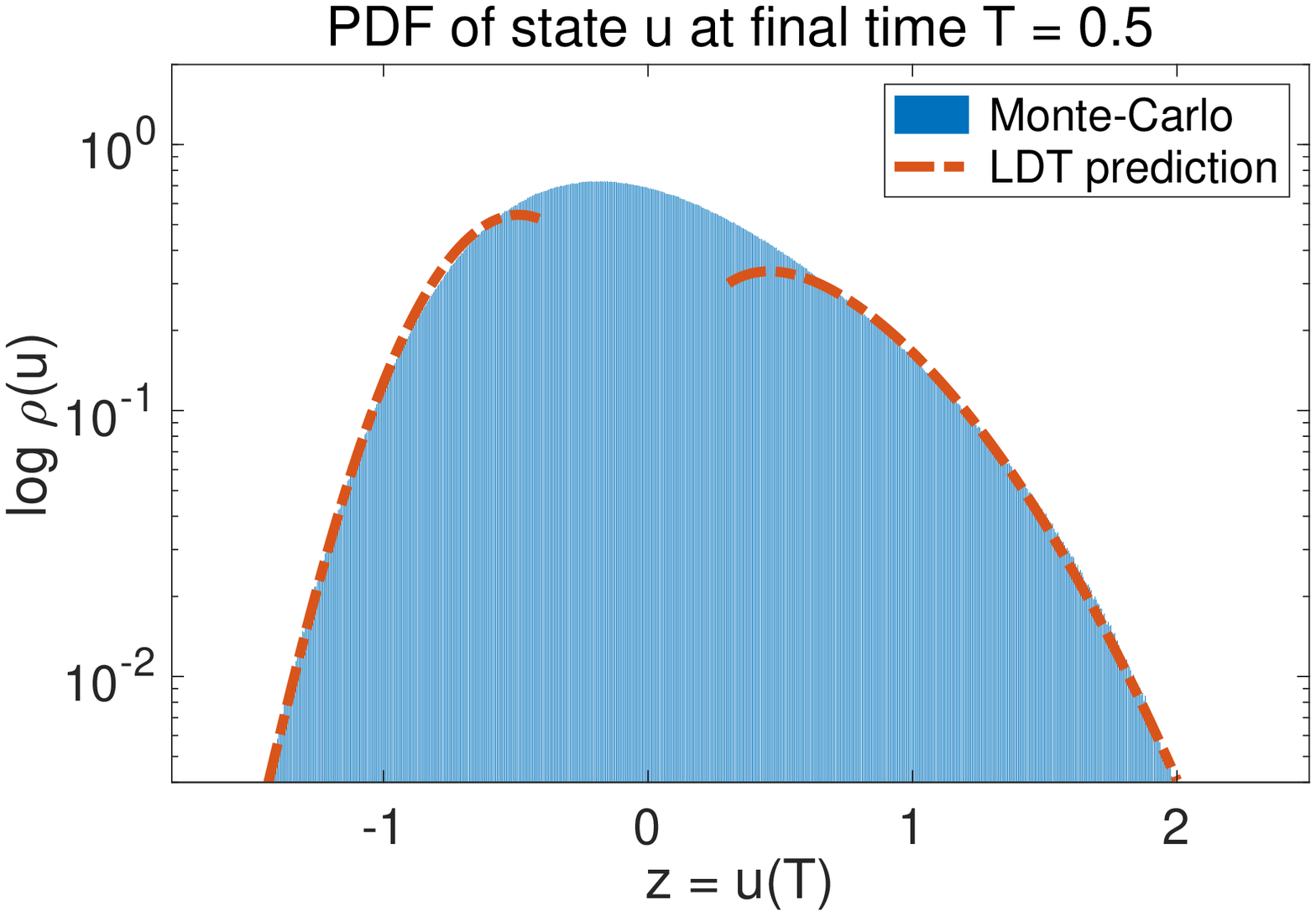}\includegraphics[scale=0.29]{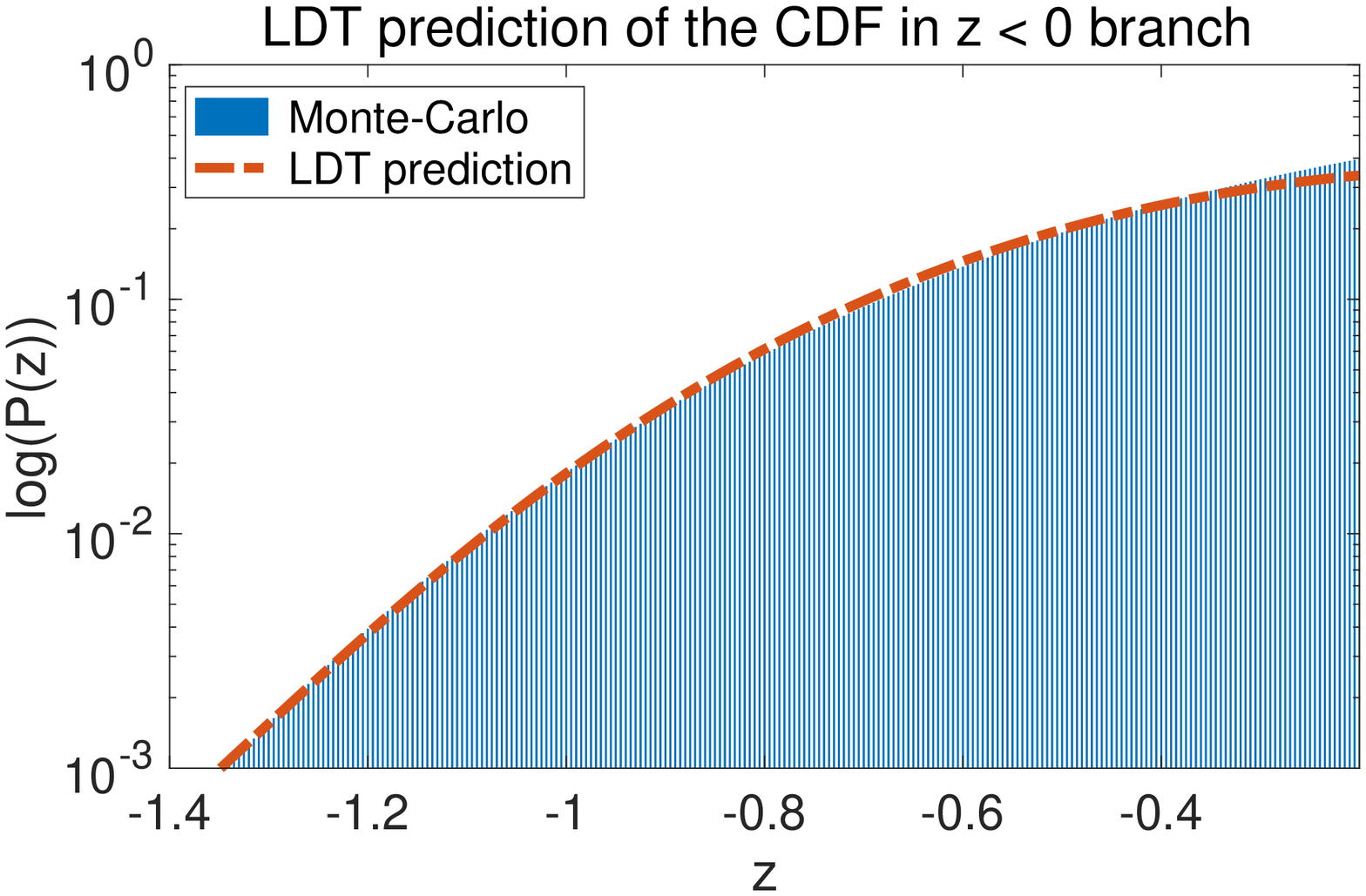}\includegraphics[scale=0.29]{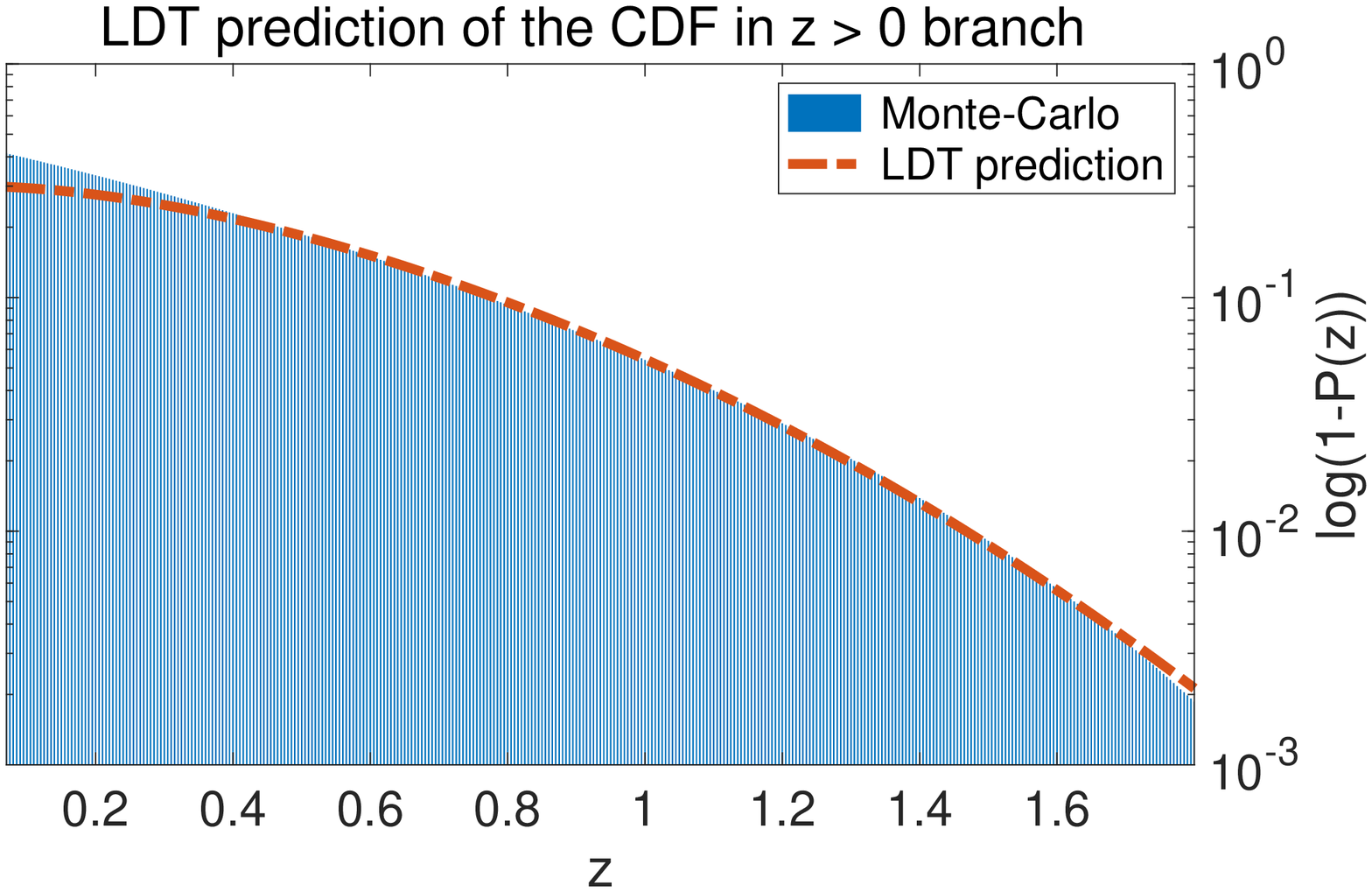}

\caption{LDT prediction of the CDF and PDF of the TKdV solution $u_{\Lambda}$
measured at $x_{c}=0$ and $T=0.5$ starting from the Gibbs invariant
measure.\label{fig:LDT-prediction-pdf}}
\end{figure}

Next, we check the ability of the LDT approach to capture the skewed shapes in the tails of the CDFs and PDFs of the solution $u_{\Lambda}$. We see  Gibbs initial state leads stronger non-Gaussian final states (see the PDFs in Figure \ref{fig:Comparison-of-PDFs} and compare with the Gaussian parameter case with a weaker skewness).
In Figure \ref{fig:LDT-prediction-pdf} we compare the LDT prediction of the CDF and PDF with those obtained by direct simulation. From the histograms from Monte-Carlo samples, we observe the stronger preference in positive extreme values whose signature is a longer tail with slower slope compared with the sharper tail in the negative branch of extreme events. This characterizes the key feature in the downstream surface water waves after the ADC consistent with the experiment observations \cite{bolles2019anomalous}.
The LDT prediction captures the shapes of the PDF and CDF tails up to even moderate to small values of $z$. In addition, the asymmetric skewed structure in the positive and negative sides of the solutions is also predicted with accuracy by the LDT approach.
It successfully dictates the probability of extreme events consistent with the very expensive direct Monte-Carlo results. Notice that the computation
of the PDF requires taking the derivative of the CDF and is achieved by the least square fitting
of the discrete data using polynomials (see the Appendix). This leads to large numerical
errors as the value of $z$ decreases. Thus the accuracy in the PDF predicted by the LDT approach deteriorates when $z$ decreases to small values. It would be interesting to improve these predictions by calculating  a higher-order correction to the probability distributions in this non-Gaussian case similar to the prefactor in the Gaussian case.

\subsection{LDT extreme solutions compared with direct simulations}

\begin{figure}
\subfloat{\includegraphics[scale=0.36]{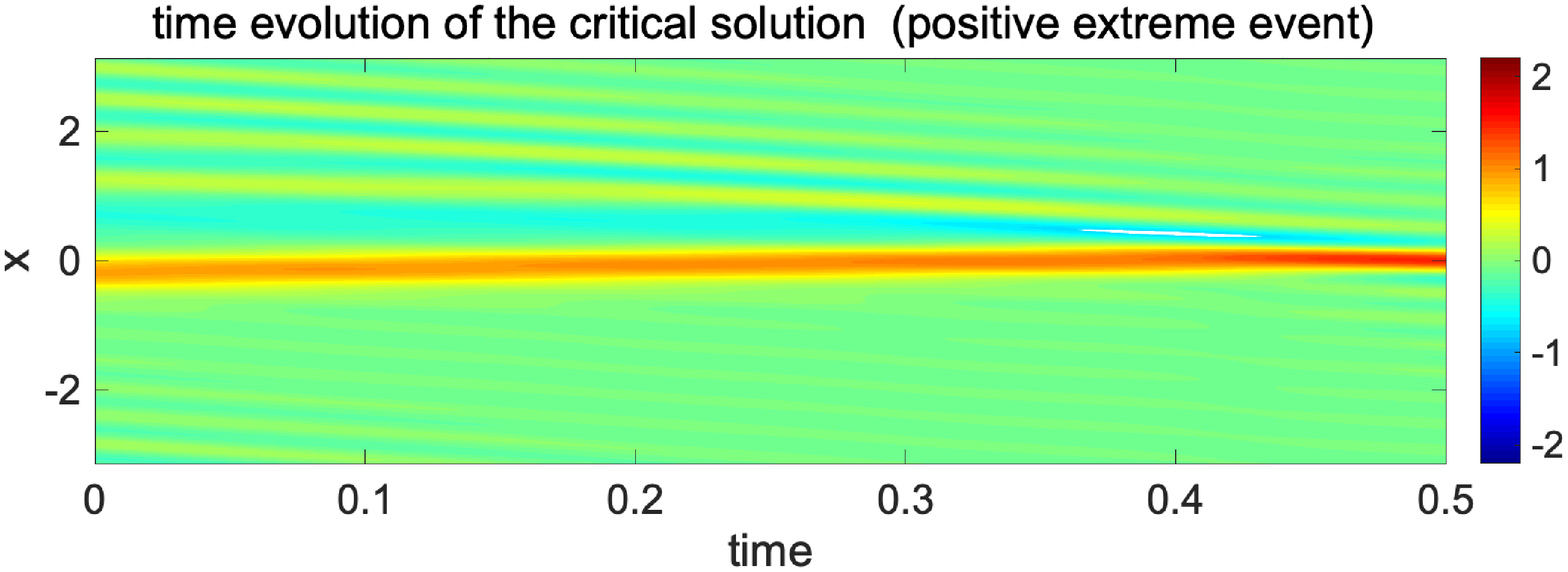}\includegraphics[scale=0.36]{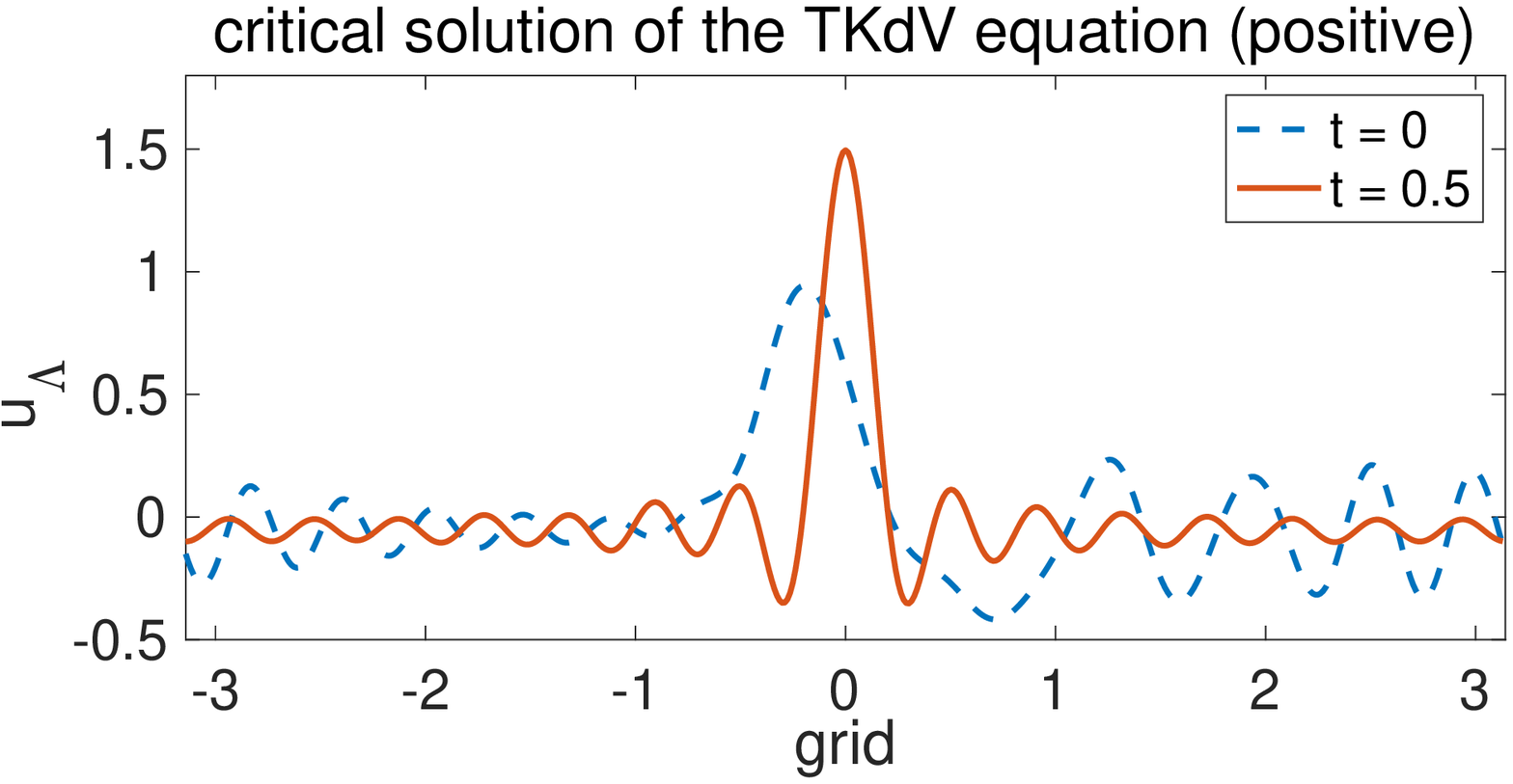}}

\subfloat{\includegraphics[scale=0.36]{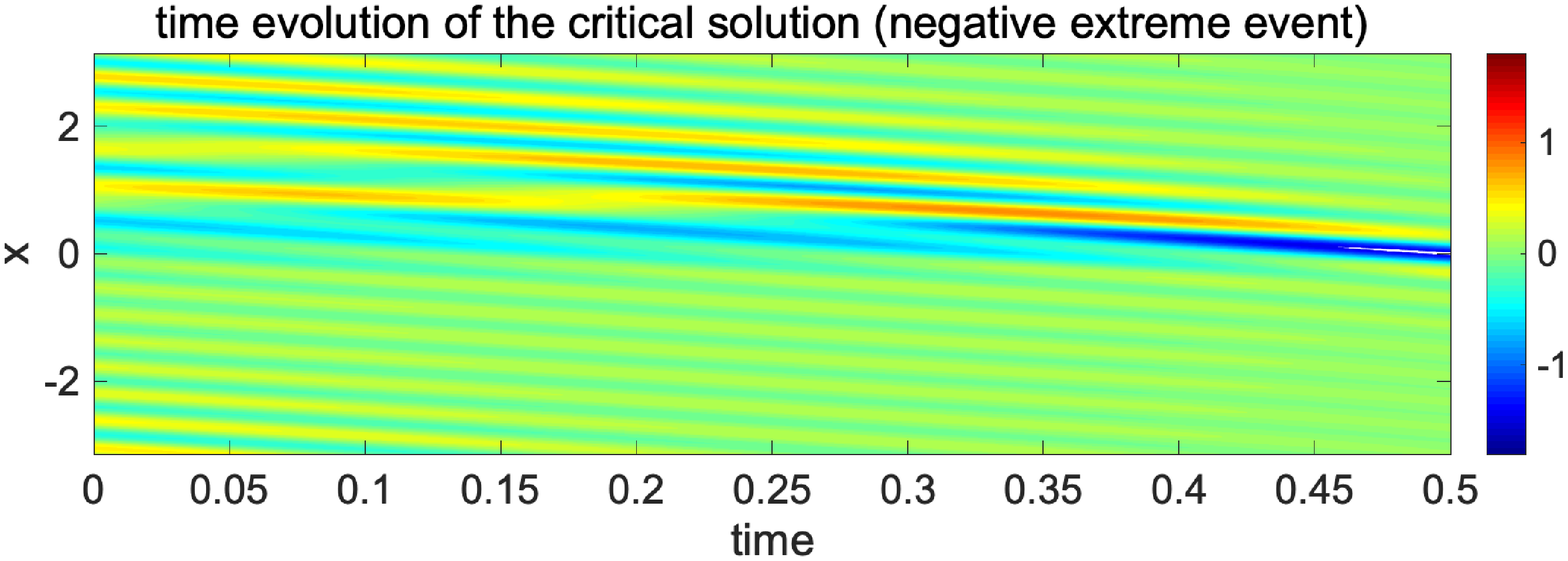}\includegraphics[scale=0.36]{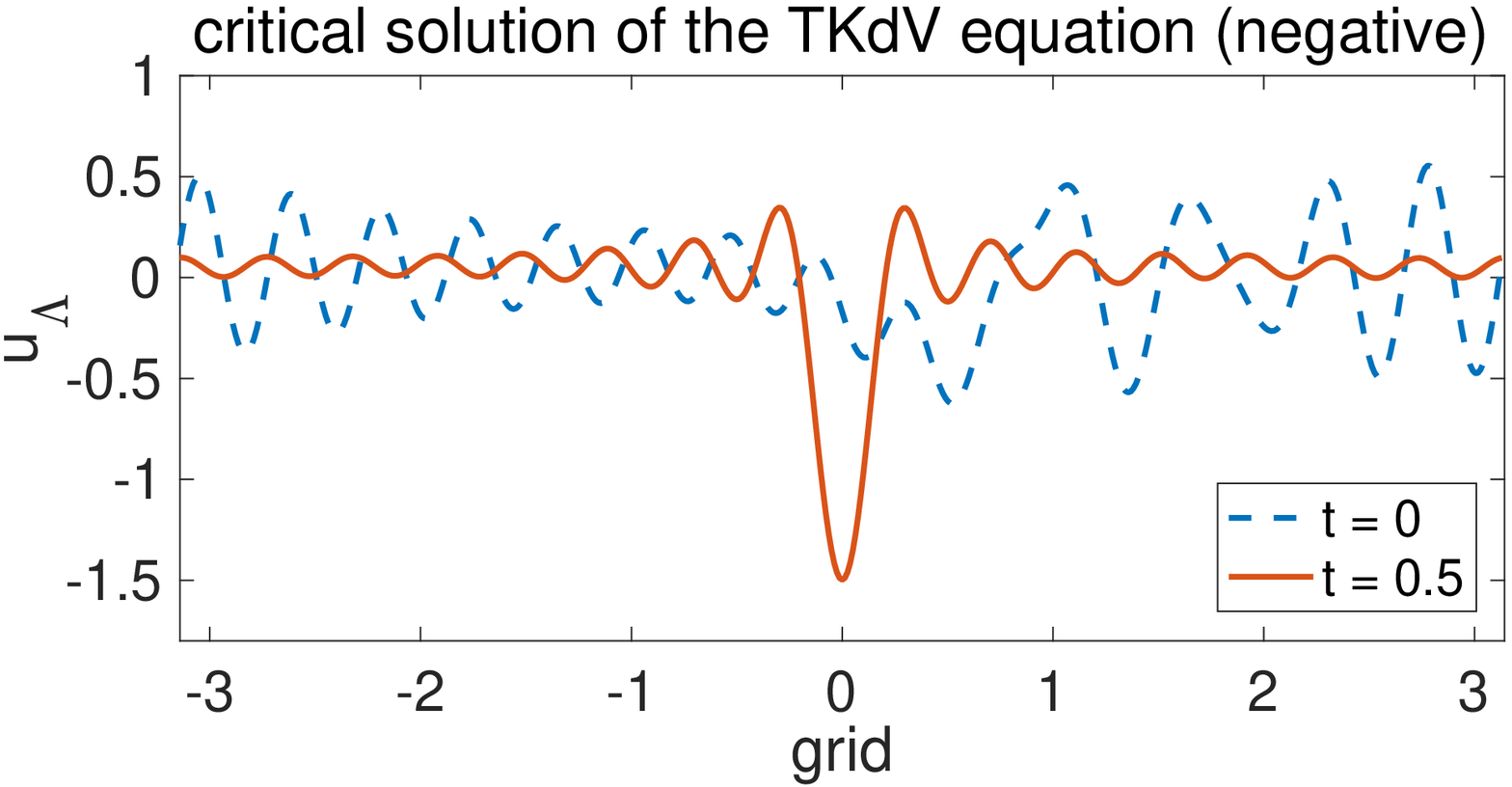}}

\caption{Time evolution of the LDT trajectory of the TKdV solution starting from Gibbs initial state to positive and negative extreme events occurring at $T=0.5,x_{c}=0$.\label{fig:trajectory-critical}}
\end{figure}

\begin{figure}

\subfloat[development of positive extreme value at different time instants]{\includegraphics[scale=0.34]{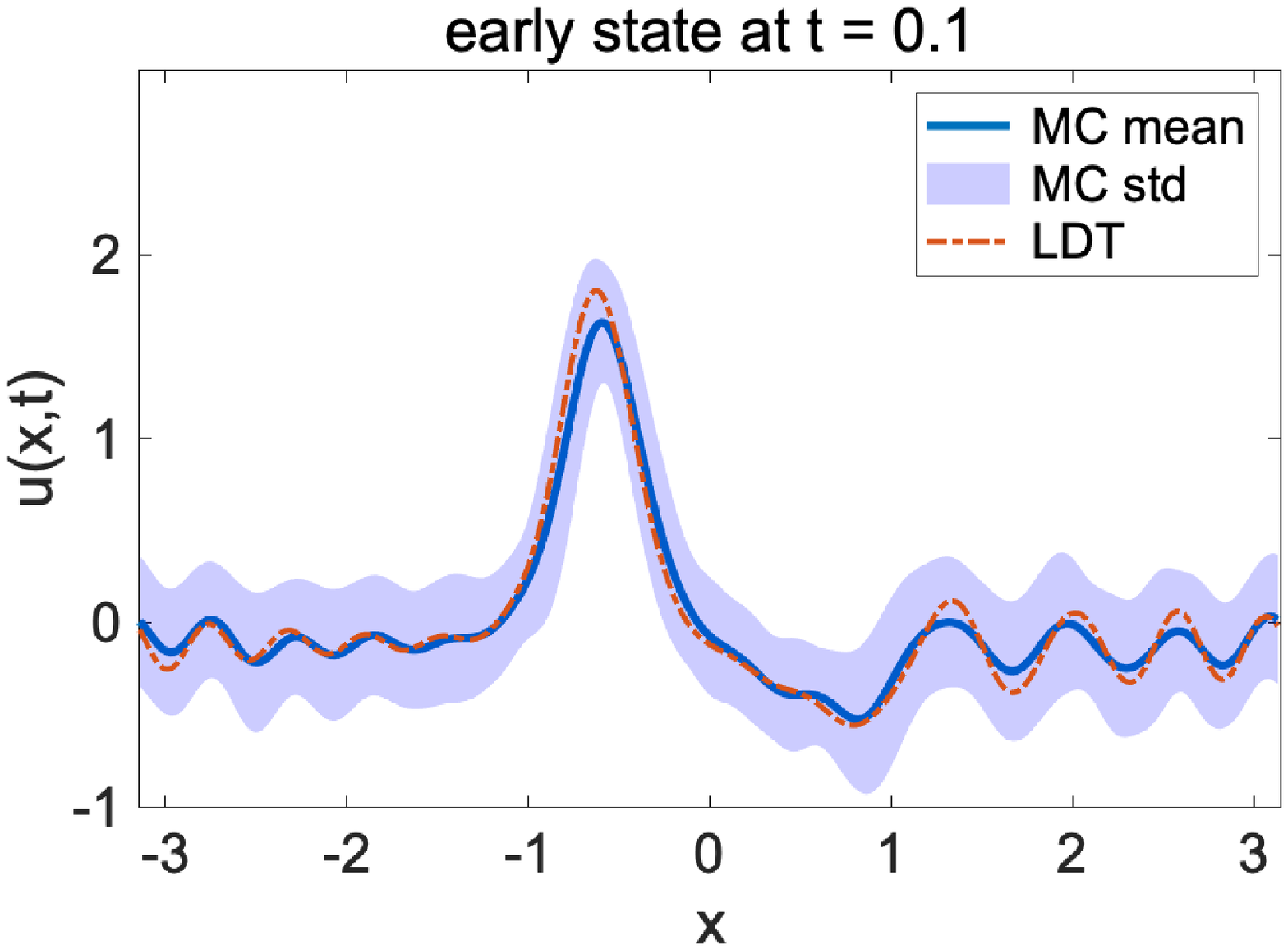}\includegraphics[scale=0.34]{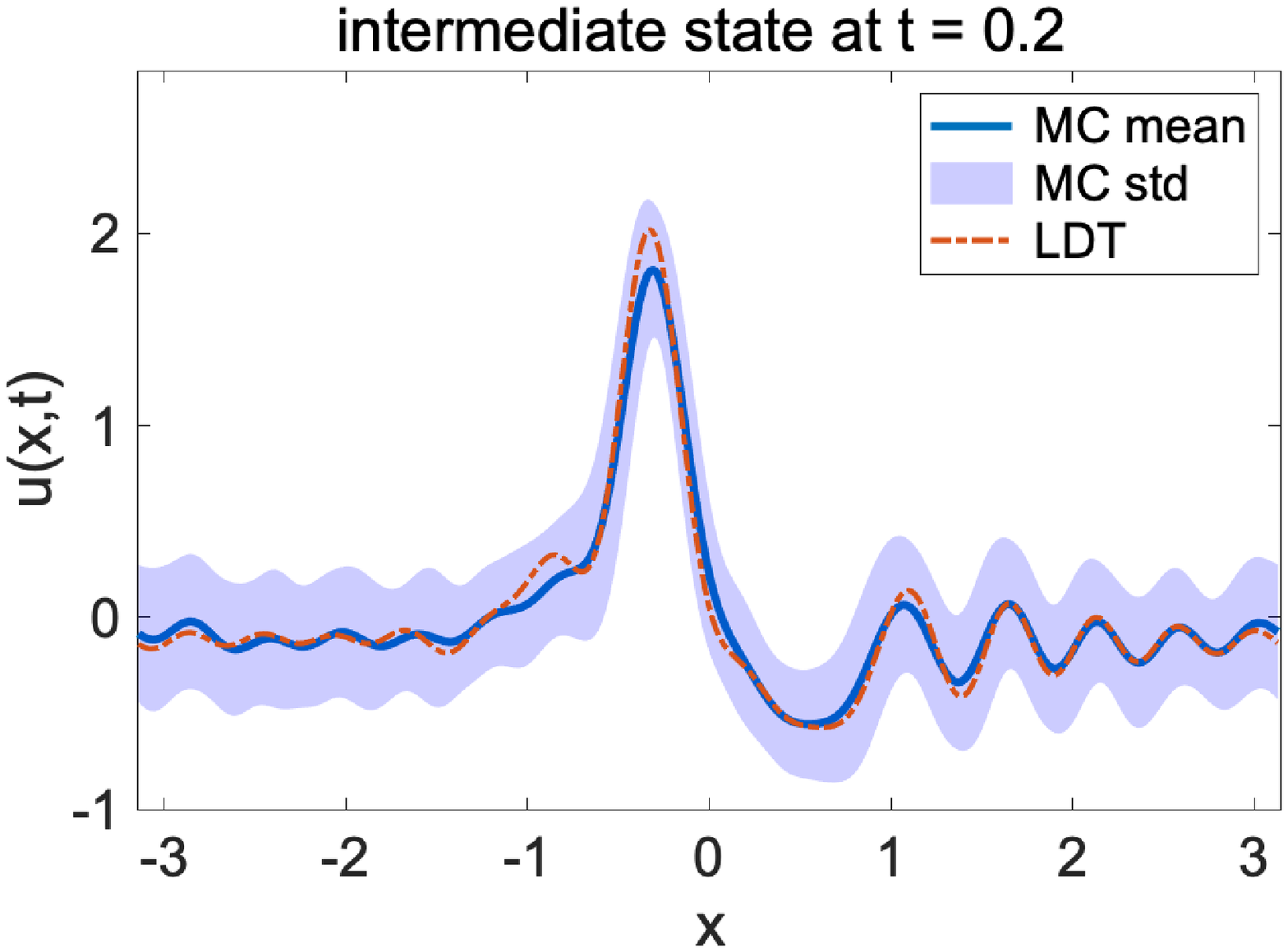}\includegraphics[scale=0.34]{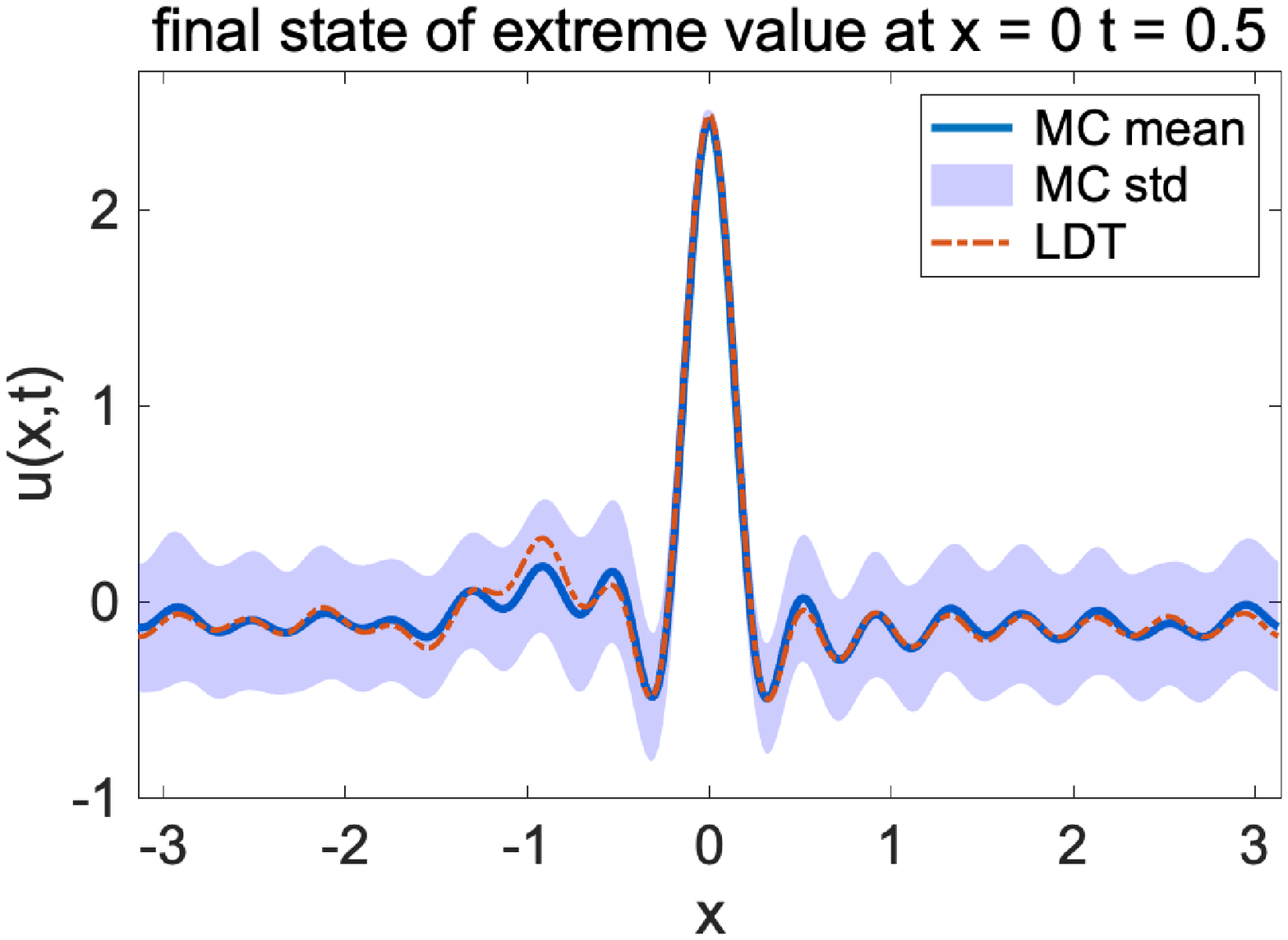}
}

\subfloat[development of negative extreme value at different time instants]{\includegraphics[scale=0.34]{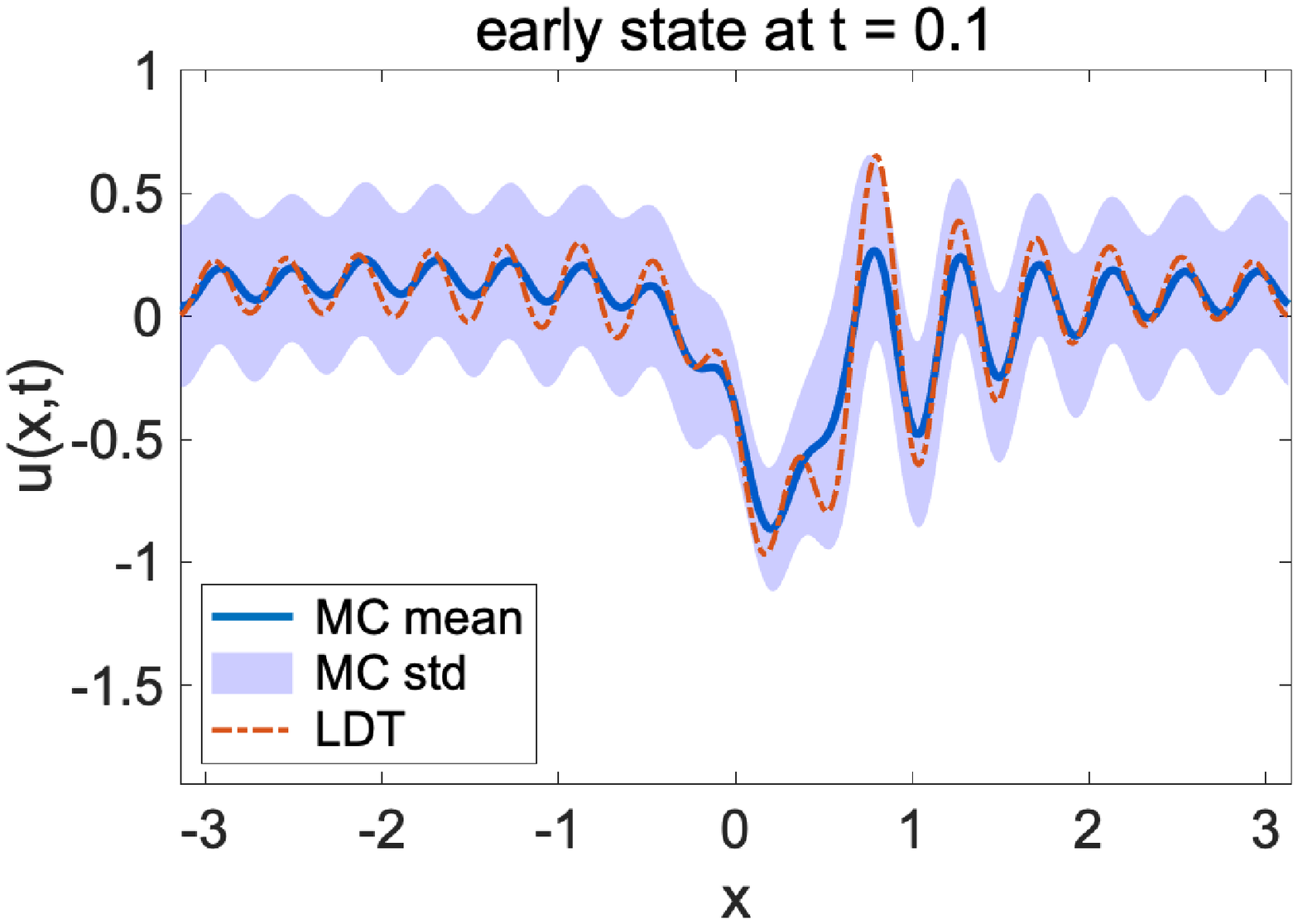}\includegraphics[scale=0.34]{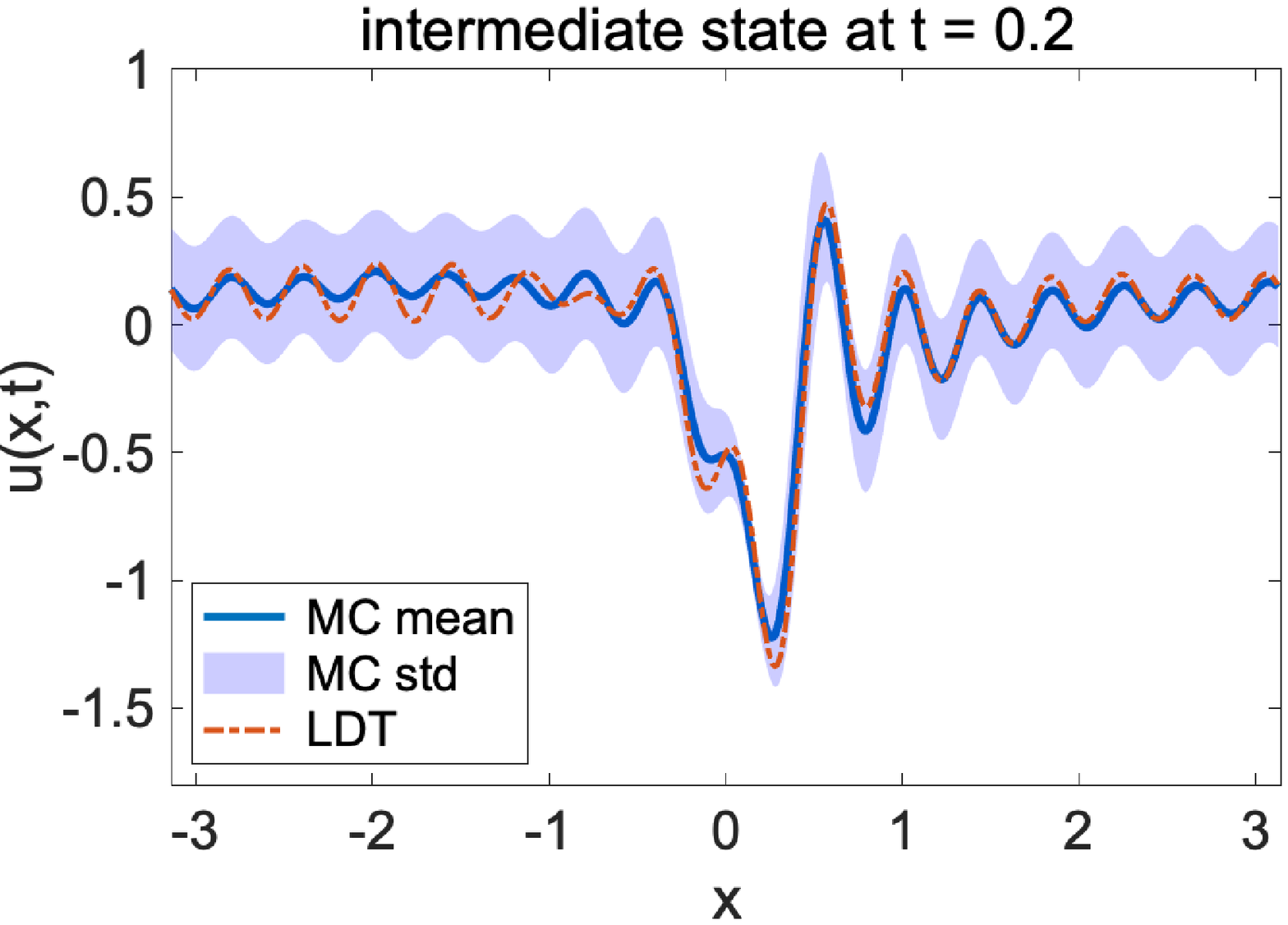}\includegraphics[scale=0.34]{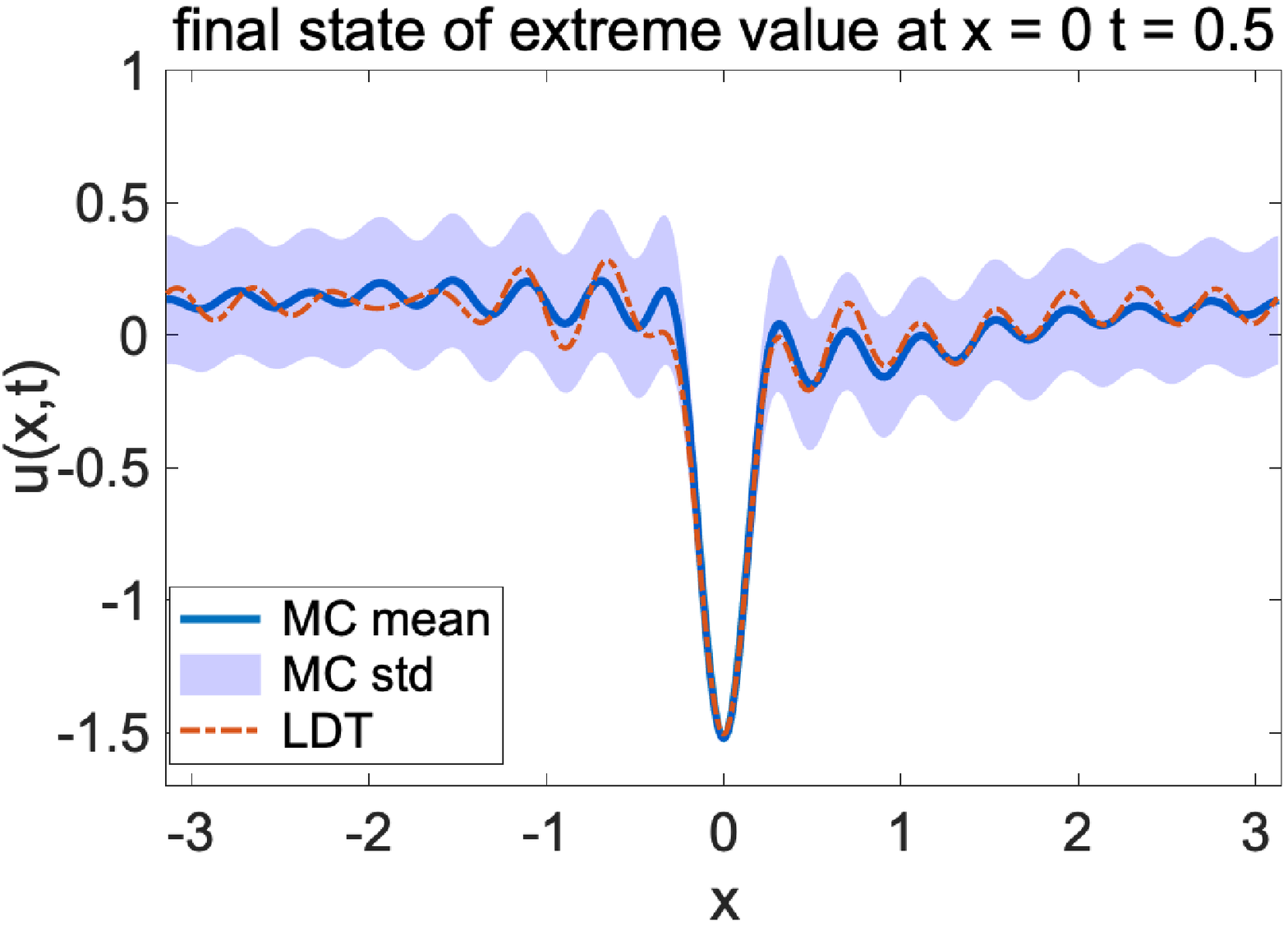}
}

\caption{Comparison between the LDT solution and the direct Monte-Carlo
solution with positive and negative extreme values. The solutions
are compared at initial ($t=0.1$) and final ($t=0.5$) time as well
as an intermediate time at $t=0.2$. The MC samples starts from the
initial Gibbs measure and extreme trajectories that reach
the extreme values are collected. One standard deviation $\pm\sigma$ of the MC samples is plotted as the shaded area around the sample mean.\label{fig:Comparison-critical}}
\end{figure}

Finally, we take a closer look at the LDT trajectories that develop into the extreme events at the final measured time. Figure \ref{fig:trajectory-critical} shows the route to extreme events from Gibbs initial data predicted by the LDT solutions. In Sec.~\ref{sec:Strategies-for-predicting} and Ref. \cite{dematteis2019extreme}, it is shown that the ensemble mean of the Monte-Carlo samples will converge to the most likely LDT solution, while the ensemble variance characterizes the error in the probability estimation. Here we use numerical results to confirm that LDT can indeed recover the solution that is mostly likely observed in direct simulations. The time evolutions of the two representative trajectories with positive and negative extreme events are plotted in the upper and lower rows of Figure \ref{fig:Comparison-critical} respectively.
We compare the solution from the LDT approach with the corresponding ensemble mean and variance from the direct simulation with the same extreme event value reached at $T=0.5, x_{c}=0$. The profiles of the LDT solution at several different time instants during the time evolution are compared with the statistical mean in the Monte-Carlo solutions with extreme events. One standard deviation of ensemble samples is also plotted in shaded areas around the ensemble mean solution to illustrate the uncertainty in the extreme solutions.
Good agreement is found with the LDT result that closely tracks the Monte-Carlo solution during their entire evolution time, especially around the region with extreme values showing very little uncertainty.
In addition, we again observe the distinctive structures in the initial state for the development of the positive and negative extreme events at a later time, as previously illustrated in Figure \ref{fig:spec_comp}.
The positive extreme event always requires a dominant wave transporting in space; in contrast the negative extreme event is the result of a group of dispersive high wavenumber waves merging together to form a final large negative value.

\section{Concluding remarks}\label{sec:summary}

In this paper, we tested  a numerical strategy based on LDT to predict the time evolution of extreme events and the related skewed probability distributions in turbulent water waves going through an abrupt depth change. We showed that the development of positive and negative extreme values from different initial configurations is accurately captured via the optimization of a proper action functional provided by LDT. We tested the LDT prediction for  two types of uncertainties including a Gaussian initial state, and a non-Gaussian initial state described by Gibbs  distribution. Detailed numerical tests were performed to illustrate the ability of the LDT approach to identify the least unlikely solution leading to an extreme event, without having to run direct Monte-Carlo simulations. In the Gaussian parameter case, a high-order correction is made to extend the LDT accuracy up to small values along the asymmetric tails in probability distributions. In the more challenging non-Gaussian case, the LDT strategy maintains its ability to capture the entire time evolution to extreme events and predict highly skewed PDFs under the highly nonlinear dynamics with non-Gaussian statistics. As a continuation of this research, it would be interesting to develop precise theories as a systematic analysis of the LDT approximation with non-Gaussian parameters and nonlinear dynamics. Generalization with high-order prefactors can be also computed to improve the accuracy of the prediction in a wider class of systems admitting a Hamiltonian structure.

\section*{Acknowledgment}
The authors would like to express their gratitude to the late A. J. Majda for many fruitful discussions and inspiring comments about this work. D. Q. was partially supported by the Office of Naval Research N00014-19-1-2286. E. V.-E. was supported in part by the NSF Materials Research Science and Engineering Center Program grant DMR 1420073, by NSF grant DMS 152276, by the Simons Collaboration on Wave Turbulence, grant 617006, and by ONR grant N4551-NV-ONR. 

\appendix
\renewcommand{\theequation}{A.\arabic{equation}}
\setcounter{equation}{0}

\section*{Appendix: Numerical algorithm for solving the LDT optimization problem}\label{sec:num_scheme}

Here, we briefly summarize the numerical strategy to find the LDT 
solution through optimization \cite{grafke2019numerical}. For practical implementation of the
optimization scheme, we seek the constrained minimizer
$\xi^{*}$ of the action functional $E_{T}$ from \eqref{eq:optim_gau} or \eqref{eq:optim_gibbs} to reach the
maximum value at the prediction time $t=T$
\begin{equation}
\xi^{*}\left(\lambda\right) =\inf_{\xi}E_{T}\left(\xi;\lambda\right).\label{eq:cost_func}
\end{equation}
In the above optimization problem, the extreme value is determined
by the optimized solution $z=u\left(\xi^{*},\lambda\right)$
from the Lagrangian multiplier $\lambda$ by minimizing the action
functional $E_{T}$. Therefore, a series of different extreme values
of $z$ can be reached in the optimization problem (\ref{eq:cost_func})
by changing the value of the Lagrangian multiplier $\lambda$. Positive
values of $\lambda>0$ lead to positive extreme values, while negative
$\lambda<0$ gives the extremes in the negative side. In this way,
the statistical prediction of a stochastic process with uncertainty
is converted to the deterministic optimization problem for
the model parameter $\xi=\left[\hat{\xi}_{0},\hat{\xi}_{1}^{r},\hat{\xi}_{1}^{i},\cdots,\hat{\xi}_{\Lambda}^{r},\hat{\xi}_{\Lambda}^{i}\right]\in\mathbb{R}^{2\Lambda+1}$. The optimized solution is achieved by the following iterative algorithm via steepest descent:

\begin{algorithm*}
Pick initial parameter value $\xi^{0}$ from a standard normal
distribution. Use steepest descent method to update the parameters
\begin{equation}
\xi^{n+1}=\xi^{n}-\alpha^{n}C^{-1}\left(\nabla_{\xi}E_{T}\right)^{n}.\label{eq:updating}
\end{equation}
For each updating step:
\begin{itemize}
\item Solve the TKdV equation (\ref{eq:model}) for $\hat{u}^{n}$ starting
from the initial value $\xi^{n}$;
\item Compute the gradient $\left(\nabla_{\xi}E_{T}\right)^{n}$
using the computed solution at the final time $T$;
\item Update the model parameter with adaptive step size $\alpha^{n}$ from
proper line searching method.
\end{itemize}
The iteration is terminated when the relative error is smaller than
a tolerance, $\mathrm{err}^{n}=\left|\xi^{n+1}-\xi^{n}\right|\leq\epsilon$,
and $\mathrm{err}^{n}>\mathrm{err}^{n-1}$.
\end{algorithm*}

In \eqref{eq:updating} for simplicity, a fixed identity matrix $C=\mathrm{Id}$ is used for the pre-conditioning, and the gradient for complex variables is computed by $\nabla_{\hat{\xi}_{k}}E_{T}=\frac{\partial E_{T}}{\partial\hat{\xi}_{k}^{r}}+i\frac{\partial E_{T}}{\partial\hat{\xi}_{k}^{i}}$.
In addition, the prediction time $T$ should be within the predictable range so that the turbulent solution still remains tractable, whereas practical numerical experiments confirm a long prediction time $T$.

We can also compute the corresponding CDF and PDF $\rho\left(z\right)$ for the TKdV state $u_{\Lambda}\left(T,x_{c}\right)$ according to the refined probability distribution $P_{T}\left(z\right)$ in~\eqref{eq:expansion_gau},  and the optimal solution $\xi^{*}\left(z\right)$ from the steepest descent of (\ref{eq:updating}). The refined probability for extreme events can be computed by
\[
P_{T}\left(z\right)\sim\exp\left[-U_{T}\left(\lambda\left(z\right)\right)\right],
\]
where we define the refined rate functional $U_{T}\left(\lambda\right)=\frac{1}{2}|\xi^{*}|^{2}+\log|\xi^{*}|+c$ as a function of $\lambda$ (then equivalently $z$). The two branches of positive and negative extreme values need to be treated separately in the PDFs for large positive values and negative values 
\begin{equation}
\begin{aligned}\rho_{T}^{+}\left(z\right)= & \:-P_{T}^{\prime}\left(z\right)\sim U_{T}^{\prime}\left(z\right)P_{T}\left(z\right),\quad z\gg1;\\
\rho_{T}^{-}\left(z\right)= & \:P_{T}^{\prime}\left(z\right)\sim-U_{T}^{\prime}\left(z\right)P_{T}\left(z\right),\quad z\ll-1.
\end{aligned}
\label{eq:pdf}
\end{equation}
If we only consider the leading-order LDT approximation (\ref{eq:LDT}), the PDF can be estimated in a crude way as $\rho_{T}\left(z\right)\sim I_{T}^{\prime}\left(z\right)\exp\left(-I_{T}\left(z\right)\right)$.
The above formulas for the PDFs require the computation of derivatives
of the refined action functional $U_{T}\left(z\right)$ or $I_{T}\left(z\right)$.
In practice, we use a 5th-order polynomial to fit the discrete converged
points from the series of values of $\lambda$.

\bibliographystyle{plain}
\bibliography{refs}

\end{document}